\documentclass[eqsecnum,twocolumn,aps,tightenlines,floats,prd]{revtex4-1}
\def\bea{\begin{eqnarray}}
\def\eea{\end{eqnarray}}
\def\bal{\begin{align}}
\def\eal{\end{align}}

\def\sfrac#1#2{{\textstyle \frac{#1}{#2}}}

\usepackage{epsf} 
\usepackage{graphics}
\usepackage{graphicx}
\usepackage{epsf} 
\usepackage{amsmath}
\usepackage{amssymb}
\usepackage{bbold}
\usepackage{slashed}
\usepackage{appendix}
\usepackage[usenames,dvipsnames]{color}

\newcommand{\B}[1] {\color{RubineRed}#1}%

\begin{document}  


\phantom{0}
\hspace{5.5in}\parbox{1.5in}{ \leftline{JLAB-THY-14-1985}
                \leftline{}\leftline{}\leftline{}\leftline{}
}
\title
{\bf Covariant Spectator Theory of $np$ scattering: Deuteron Quadrupole Moment}

\author{Franz Gross$^{1,2}$ 
\vspace{-0.1in}  }

\affiliation{
$^1$Thomas Jefferson National Accelerator Facility, Newport News, VA 23606 \vspace{-0.15in}}
\affiliation{
$^2$College of William and Mary, Williamsburg, Virginia 23185}


\date{\today}

\begin{abstract} 

The deuteron quadrupole moment is calculated using two CST model wave functions obtained from the 2007 high precision fits to $np$ scattering data.  Included in the calculation are a new class of isoscalar $np$ interaction currents  automatically generated by the nuclear force model used in these fits. 
  The prediction for model WJC-1, with larger relativistic P-state components, is $2.5\%$ smaller that the experimental result, in common with the inability of models prior to 2014 to  predict this important quantity.  However, model WJC-2, with very small P-state components, gives agreement to better than 1\%,  similar to the results obtained  recently  from  $\chi$EFT predictions to order N$^3$LO.  
 
\end{abstract}
 
\phantom{0}

\maketitle


\section{Introduction and Background} 

\begin{table}[b]
\begin{minipage}{3.5in}
\caption{Predictions of the quadrupole moment, expressed as an ``error'' defined by $\delta Q_{\rm pred}=(Q_{\rm pred}-Q_{\rm exp})/Q_{\rm exp}$. 
} 
\label{tab:predictions}
\begin{ruledtabular}
\begin{tabular}{ll}
Reference  & $\quad \delta Q_{\rm pred}$    (model) \cr
\tableline  
GVOH  \cite{Gross:1991pm} & $-9.0\%$ (IIB),  $-8.1\%$ (IIB with RC) \cr
Argonne 
\cite{Wiringa:1994wb} & $-3.8\%$ (with MEC)  \cr 
CD Bonn \cite{Machleidt:2000ge} & $-5.6\%$ (no MEC), $-2.1\%$ (MEC est.) \cr
Light Front 
\cite{Huang:2008jd} &  $-5.7\%$ (IM), $-3.8\%$ (IM+Ex)\cr
\tableline
$\chi$EFT (ODU-Pisa) \cite{Piarulli:2012bn} & $-0.3\%$ (500), $-1.4\%$ (600)\cr
this work & $-2.5\%$ (WJC-1), $-0.8\%$ (WJC-2)
\end{tabular}
\end{ruledtabular}
\end{minipage}
\end{table}

\begin{figure*}
\centerline{
\mbox{
\includegraphics[width=7in]{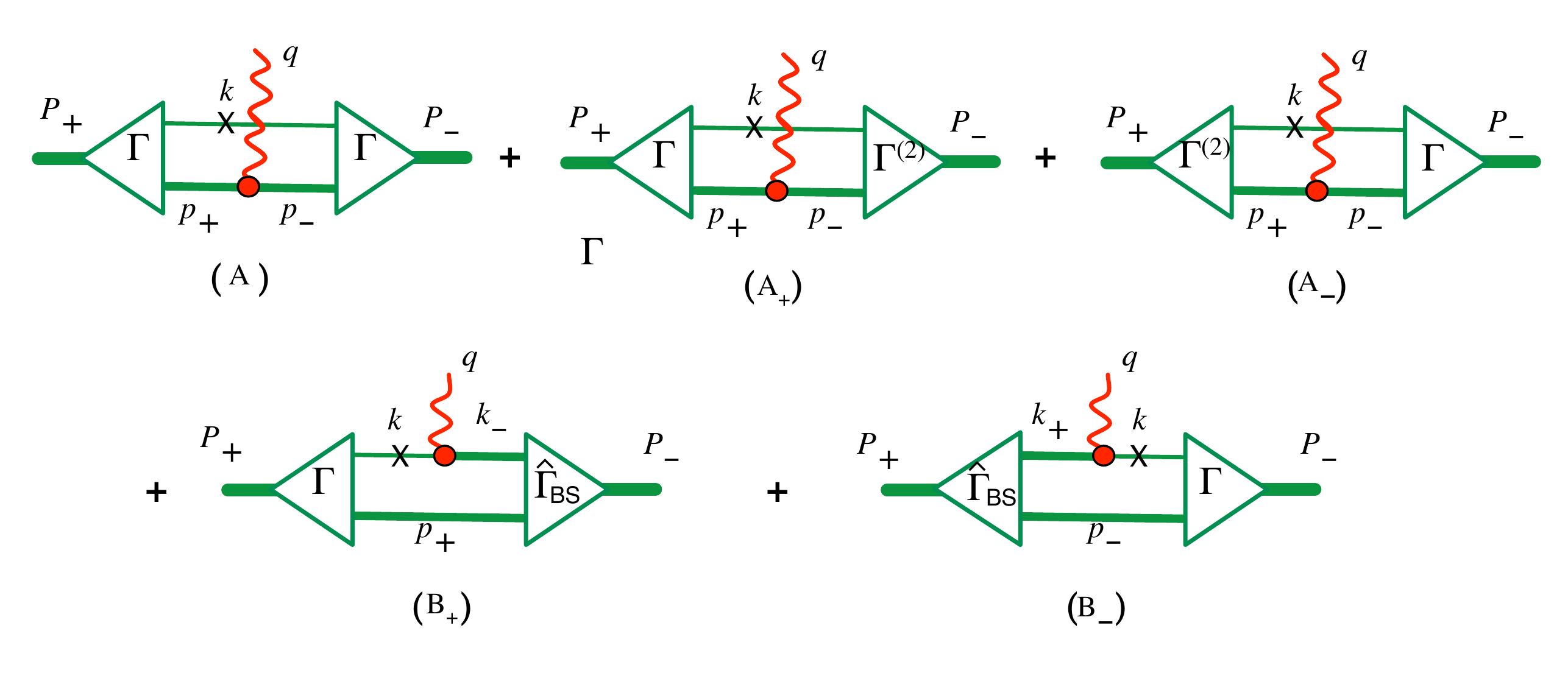}
}
}
\caption{\footnotesize\baselineskip=10pt (Color on line) Diagrammatic representation of the two-body current operator in the Covariant Spectator Theory using vertex functions with particle 2 off-shell.  The interaction current contributions are contained in diagrams (A$_\pm$) and parts of the (B) diagrams, as discussed in the text. Off-shell nucleon lines are thicker than on-shell lines, which are also labeled with an $\times$.  Diagrams (A) and (A$_\pm$) describe the interaction of the photon with particle 2, allowing particle 1 to be on-shell in both the initial and final state.  Diagrams (B$_\pm$)  describe the interaction of the photon with particle 1, so that  both particles must off-shell in either the initial state (diagram B$_+$) or in the final state (diagram B$_-$).
}
\label{Fig1}
\end{figure*} 

Until recently, calculations of the deuteron quadrupole moment consistently under-predicted its value by several percent \cite{Gross:1991pm,Wiringa:1994wb,Machleidt:2000ge,Huang:2008jd}.  Some of these calculations are summarized in Table \ref{tab:predictions}; all of the results shown there use realistic NN scattering models with kernels or potentials adjusted to fit the low entergy NN data.  Because of these fits, predictions of the quadruple moment are very tightly constrained, with uncertainties coming only from relativistic  corrections, including those to the current operator, which have been difficult to determine.  The difficulty of avoiding these constraints led Machleidt \cite{Machleidt:2000ge} to identified the under prediction of the quadrupole moment as  an ``unresolved problem.''  

Now, a new chiral effective field theory ($\chi$EFT) calculation, done to order N$^3$LO  by the ODU-Pisa group \cite{Piarulli:2012bn},  has obtained very good agreement.  Two unknown isoscalar low energy constants (LEC's) appear to this order, and the ODU-Pisa group fixes them  by fitting  the deuteron  and the isoscalar trinucleon magnetic moments.  However, all of their results still depend on the cutoff $\Lambda$, which is needed to renormalize the calculations,  and the dependence of their result for the quadrupole moment on $\Lambda$ (for the two values of 500 and 600 MeV studied) is shown in Table \ref{tab:predictions}.  The dependence on $\Lambda$ is not strong, and both results are much closer to the experimental result that found previously.

The principal purpose of this paper is to report the new results for the quadrupole moment obtained from the two high precision models of NN scattering (WJC-1 and WJC-2) that Stadler and I found in 2008 \cite{Gross:2008ps} using the covariant spectator theory (CST) \cite{Gross:1969rv,Gross:1972ye,Gross:1982nz}.  (The features and differences between these two models will be briefly reviewed in Sec.~\ref{sec:imp}.)   The predictions obtained for these two models are shown in the last line of Table \ref{tab:predictions}.  The physics that went into these calculations of the quadrupole moment (which contain {\it no free parameters\/}) will be very briefly summarized in Sec. \ref{sec:cal}, with all of the extensive details moved to the Appendix.  The implications of these  results are discussed in Sec.~\ref{sec:imp}.

This paper is the third in a series of four planned papers.  These papers grew out of the need for a new treatment of the NN current required by the nature of the kernels used in the  high precision fits of 2008.  At that time it was  found that a kernel consisting of a sum of covariant  one-boson exchange (OBE) diagrams would  give an excellent high-precision fit to the NN data (with a $\chi^2$/datum $\simeq1$) provided that the vertex function, $\Lambda^{\sigma_0}$, that describes the coupling of the scalar-isoscalar boson $\sigma_0$ to the nucleon,  included momentum-dependent terms of the form
\bea
\Lambda^{\sigma_0}(p,p')=g_{\sigma_0}{\bf 1} -\nu_s[\Theta(p)+\Theta(p')] \label{eq:1pt1}
\eea
where $g_s$ and $\nu_s$ are parameters adjusted to fit the data, $p$ and $p'$ are the four-momentum of the outgoing  and incoming nucleons, and the operator $\Theta$ is the negative-energy projection operator for a spin 1/2 nucleon
\bea
\Theta(p)=\frac{m-\slashed{p}}{2m} \, .  \label{eq:1pt2}
\eea
For more discussion of these OBE models, see Ref.~\cite{Gross:2008ps}.

These $\nu$ dependent terms, which vanish when the nucleons are on-shell, introduce a new kind of energy dependence into the kernel, generating a new class of isoscalar interaction currents.  The first  paper in this series, referred to as Ref.~I \cite{Gross:2014zra}, showed how current conservation \cite{Gross:1987bu} and principles of picture independence and simplicity could be used to {\it uniquely\/} determine these interaction currents.  Then, in the second paper, referred to as Ref.~II \cite{Gross:2014wqa}, I calculated the deuteron magnetic moment and showed that both  high precision models gave a nearly identical prediction that is only about 1\% larger than the experimental value.  The magnetic moment cannot distinguish between the two models.  However the predictions for the quadruple moment shown in Table \ref{tab:predictions} provide a basis for distinguishing between the two models and this will be discussed in  Sec.~\ref{sec:imp}.

\section{summary of the calculation} \label{sec:cal}

\begin{figure*}
\centerline{
\mbox{
\includegraphics[width=6.5in]{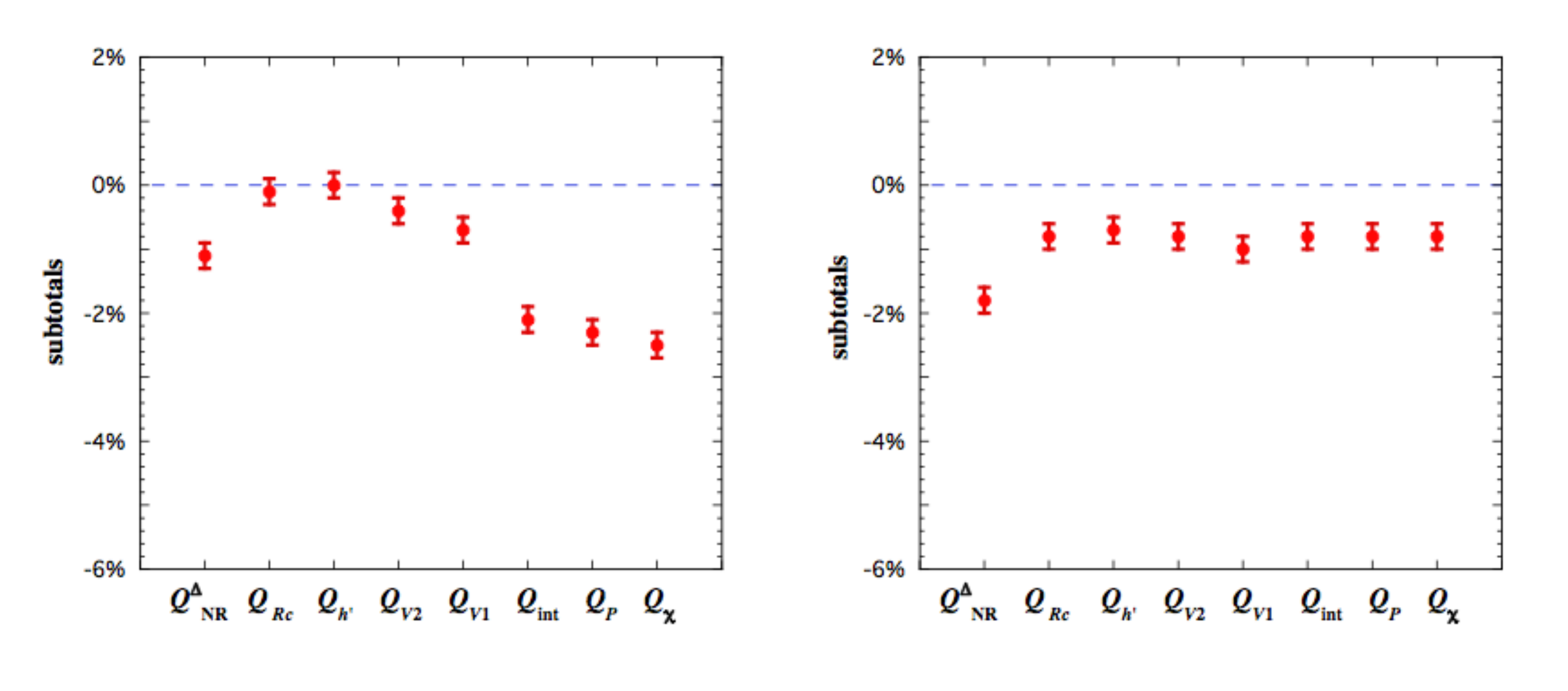}
}
}
\caption{\footnotesize\baselineskip=10pt (Color on line) Running sum of the corrections (in \%) to the quadrupole moment, in the order that they are listed in Tables \ref{tab:Qnames} and \ref{tab:Qnbrs}.  The dashed line is the experimental value (zero correction).  The error bars are $\pm0.002=\pm0.2\%$, an estimate of the size of the terms missing from the approximation of Eq.~(\ref{eq:result}).  Model WJC-1 (left panel) and Model WJC-2 (right panel).  
}
\label{fig:two}
\end{figure*} 

\begin{table}[b]
\begin{minipage}{2.8in}
\caption{Physical origin of the eight different types of terms that contribute to the quadrupole moment. } 
\label{tab:Qnames}
\begin{ruledtabular}
\begin{tabular}{c|l}
term  & physical origin   \cr
\tableline  
$Q_{\rm NR}$ & nonrelativistic contribution from the S, D-states   \cr
 $Q_{\it Rc}$ & relativistic corrections to S,D terms \cr
 $Q_{h'}$ & dependence on the strong form factor, $h$ $\quad$ \cr
 $Q_{V_2}$ & interaction currents: off-shell particle 2 \cr
 $Q_{V_1}$ &  interaction currents: on-shell particle 1 \cr
$Q_{\rm int}$  &  S,D and P-state interference  \cr
 $Q_P$  &P-state squared terms  \cr 
$Q_{\chi}$  & P-state and  negative $\rho$-spin $z_\ell^{--}$ interference \cr
\end{tabular}
\end{ruledtabular}
\end{minipage}
\end{table}
%

In the CST, the two body current is given by the five diagrams shown in Fig.~\ref{Fig1} (
also shown in Ref.~II).  These include the interaction current contributions derived in Ref.~I, expressed in terms of the the effective wave functions ${\Psi}^{(2)}$ and the subtracted  vertex functions $\widehat \Gamma$ (directly related to $\widehat \Psi$) with two particles off shell.   These contributions are discussed below, but for a complete discussion of the physics, see Refs.~I and II. 

 The quadrupole moment, $Q_d$, in units of $e/m_d^2$, is extracted by taking the $Q^2\to0$ limit of the difference of two matrix elements of the two body current, ${\cal J}_i$,
\bea
Q_d=\lim_{Q^2\to0}\frac{m_d}{Q^2}\Big[{\cal J}_1-{\cal J}_2\Big] \label{eq:quad}
\eea
where the current matrix elements are 
\bea
&& {\cal J}_1\equiv G^0_{00}=2D_0\left(G_C+\frac{4}{3}\eta\, G_Q\right)\nonumber\\
&& {\cal J}_2\equiv G^0_{+-}=2D_0\left(G_C-\frac{2}{3}\eta \,G_Q\right)\, , \label{eq:dffmatrix}
\eea
with $G^{\lambda_\gamma}_{\lambda\lambda'}$  the matrix element  for an incoming (outgoing) deuteron with four-moments $P_- (P_+)$ and helicity $\lambda'  (\lambda)$ and a virtual photon with helicity $\lambda_\gamma$  
\bea
G^{\lambda_\gamma}_{\lambda\lambda'}\equiv
\left<P_+\,\lambda\right|J_\mu\left|P_-\,\lambda'\right>\epsilon_{\lambda_\gamma}^\mu \, . \label{eq:current1}
\eea
Eq,~(\ref{eq:dffmatrix}) has been evaluated in the Breit frame, where  the photon four-momentum is $q=\{0,{\bf q}\}$, and $Q^2={\bf q}^2$, $P_\pm=(D_0,\mp\frac12{\bf q})$, $D_0=\sqrt{m_d^2+Q^2/4}$.   Details can be found in Ref.~II.

The calculation of the quadrupole moment is described in the Appendix.  The final result can be arranged into a sum of the eight terms summarized in Table \ref{tab:Qnames} and given explicitly in Eq.~(\ref{eq:result}).   To understand the origin of these terms, recall that the relativistic deuteron wave function with one particle on-shell (and the other off-shell) can be expanded in terms of four relativistic wave functions: $u$ (S-state), $w$ (D-state), $v_t$ (a P-state wave function with spin triplet structure), and $v_s$ (a P-state wave function with a spin singlet structure) \cite{Buck:1979ff,Gross:2010qm}.  When both particles are off-shell, an additional four wave functions could contribute, but only one combination, the $z_\delta$ defined in Eq.~(\ref{eq:zdelta}),  contributes in leading order.  The eight terms can now be described.

\begin{table}[b]
\begin{minipage}{3.2in}
\caption{Contributions to the quadrupole moment from the eight different types of corrections discussed in the text.  All terms are normalized by the experimental value of the quadruple moment ($Q_{\rm exp}=0.286$), with $Q^\Delta_{\rm NR}=(Q_{\rm NR}-Q_{\rm exp})/Q_{\rm exp}$, so that all of these terms must sum to zero to get the correct experimental value.  } 
\label{tab:Qnbrs}
\begin{ruledtabular}
\begin{tabular}{lrrrr}
& \multicolumn{2}{c} {WJC-1} &  
\multicolumn{2}{c} {WJC-2} \cr
\tableline 
 &  $u,w$ only  & ${\rm all}\quad$ &  $u,w$ only  & ${\rm all}\quad$   \cr
\tableline  
$Q^\Delta_{\rm NR}$  & $-0.011$ & $-0.011$  &$-0.018$ & $-0.018$    \cr
 $Q_{Rc}$ & 0.010 & 0.010 & 0.010 & 0.010 \cr
 $Q_{h'}$ & 0.001 &  0.001  & 0.001 &  0.001 \cr
 $Q_{V_2}$  & $-0.004$ & $-0.004$  & $-0.001$& $-0.001$\cr
 $Q_{V_1}$ & $-0.004$ & $-0.003$ & $-0.001$& $-0.002$ \cr
$Q_{\rm int}$  & --- & $-0.014$ & --- & 0.002    \cr
 $Q_P$  & ---& $-0.002$ & ---& 0.000  \cr 
$Q_{\chi}$  & --- & $-0.002$ & --- & $-0.000$ \\[0.02in]
\tableline
total & $-0.008$ & $-0.025$ & $-0.009$ & $-0.008$
\end{tabular}
\end{ruledtabular}
\end{minipage}
\end{table}
%

The largest contribution, $Q_{\rm NR}$,  is familiar from the first days of nuclear physics  \cite{Jankus}
\bea
Q_{\rm NR}=\frac{\sqrt{2}}{10}\int_0^\infty r^2dr\Big\{uw-\frac{w^2}{\sqrt{8}}\Big\}\, . \label{eq:QNR}
\eea
However, while the same formula (\ref{eq:QNR})  arrises in both the nonrelativistic theory and (as the leading contribution) in the CST theory the two results are numerically very different because the normalization of  the $u$ and $w$ wave functions in the two cases is very different.  In the nonrelativistic theory,  the normalization is
\bea
\int_0^\infty k^2dk(u^2+w^2)&=&1
\label{eq:normNR}
\eea 
while in the CST theory it is 
\bea
\int_0^\infty k^2dk(u^2+w^2)&=&1+N_{\rm CST}
\label{eq:norm}
\eea 
 where
\bea
N_{\rm CST}&=&-\left<\frac{\partial V}{\partial m_d}\right> -\int_0^\infty k^2dk(v_t^2+v_s^2)
\label{eq:normN}
\eea 
with $V$ the NN kernel, including the strong nucleon form factors $h$, and the derivative with respect to the deuteron mass (or, alternatively, the total energy in the deuteron rest system) is a consequence of the interaction current, as discussed ion Ref.~II.  The contributions to $N_{\rm CST}$, discussed in detail in Ref.~II, are summarized in Table~\ref{tab:Nnbrs}.

\begin{table}
\begin{minipage}{3.2in}
\caption{Contributions to the normalization factor $N_{\rm CST}$ from the four different types of corrections discussed in the text (extracted from Tables I and II of Ref.~II).  } 
\label{tab:Nnbrs}
\begin{ruledtabular}
\begin{tabular}{lrrrr}
& \multicolumn{2}{c} {WJC-1} &  
\multicolumn{2}{c} {WJC-2} \cr
\tableline 
 &  $u,w$ only  & ${\rm all}\quad$ &  $u,w$ only  & ${\rm all}\quad$   \cr
\tableline  
 $N_{h'}$ & $-0.036$ &  $-0.025$  & $-0.018$ & $-0.018$ \cr
 $N_{V_2}$  & $0.022$ & $0.023$  & $0.011$& $0.011$\cr
 $N_{V_1}$ & $0.052$ & $0.057$ & $0.032$& $0.030$ \cr
 $N_P$  & ---& $-0.003$ & ---& 0.000  \cr 
\tableline
total & $0.038$ & $0.052$ & $0.025$ & $0.023$
\end{tabular}
\end{ruledtabular}
\end{minipage}
\end{table}

Hence the $Q_{\rm NR}$ of Eq.~(\ref{eq:QNR}) is larger than the nonrelativistic result by a factor of $N_{\rm CST}$ but this correction is ``hidden'' in the sense that it is already included in the leading term $Q^\Delta_{\rm NR}$ given in Table \ref{tab:Qnbrs}.  One may infer from $Q^\Delta_{\rm NR}$ that using the (incorrect) nonrelativistic normalization would give a result for the quadrupole moment about $6\%$ too small for WJC-1 and $4\%$ too small for WJC-2.

While the relativistic normalization (\ref{eq:norm}) makes a significant contribution, the calculation is not complete and the result believable until all of the other effects that come from the relativistic structure of the  interaction current and the deuteron wave functions are also calculated.   Each of these remaining effects, in the order listed in Table~\ref{tab:Qnames}, will be discussed briefly.
As in Ref.~II,  only the leading contributions to these corrections (those believed to be larger than 0.001) are retained.   A detailed discussion of which terms can be expected to be ``leading'' was presented in Ref.~II, and the same guidelines are followed here.

The $Q_{Rc}$ term includes the corrections of order $k^2/m^2$ coming from the expansion of the relativistic kinetic energy, $E_k=\sqrt{m^2+{\bf k}^2}$, which appears in many places through the calculation.  Only corrections to products involving the largest wave functions ($u$ and $w$) are leading.   This kinematical relativistic correction is one of the largest effects, and of comparable size for both models. 

The $Q_{h'}$ and $N_{h'}$ terms include corrections to the quadrupole moment that come from the strong nucleon form factor $h(p)$.  This form factor is a function of $p^2$, the four-momentum of the off-shell nucleon (only), and is normalized to unity when $p^2=m^2$.  As shown in Eq.~(\ref{eq:quad}), the calculation of the quadrupole moment requires expanding the electromagnetic  form factors around $Q^2=0$, requiring that the strong form factor be expanded around its mass-shell point, introducing  correction terms proportional to  $a(p^2)=d \log(h)/dp^2|_{p^2=m^2}$.  As shown in Table \ref{tab:Nnbrs}, terms of this type make about a  $-2\%$  contribution to the relativistic normalization already included in the leading $Q^\Delta_{\rm NR}$;  the {\it additional\/} corrections to the quadrupole moment contained in $Q_{h'}$ turn out to be negligible.

The $Q_{V_2}$ and $N_{V_2}$  terms include contributions from the isoscalar exchange current generated by the momentum dependence included in the projection operators $\Theta$ [defined in Eq.~(\ref{eq:1pt2})] that operate on the off-shell particle 2 (illustrated in the diagrams (A$_\pm$) shown in Fig.~\ref{Fig1}).  Terms of this type are present in the vertex functions for the exchange of all mesons (except the axial-vectors present in Model WJC-1), but the contributions from the pseudoscalar exchanges ($\pi$ and $\eta$) cancel.  The way in which $\Theta$ appears in the $sNN$ vertex functions for scalar ($s$) exchange was already illustrated in Eq.~(\ref{eq:1pt1}).  The structure of the exchange current implied by the appearance of these operators $\Theta$ was uniquely determined in Ref.~I where it was shown how their contributions can be expressed in terms of new deuteron wave functions generically denoted by $z^{(2)}$.  As shown in Table \ref{tab:Nnbrs}, terms of this type are already included in $Q^\Delta_{\rm NR}$, where they  make about a  $2\%$ (1\%) contribution for models WJC-1 (WJC-2);  the {\it additional\/} corrections shown in Table \ref{tab:Qnbrs} are much smaller.  

The $Q_{V_1}$ and $N_{V_1}$  terms  includes contributions from that part of the isoscalar interaction current that contributes when (the usually on-shell) particle 1 is forced of-shell by the kinematics. Explicitly,  in diagram (B$_-$) of Fig.~\ref{Fig1}, particle 1 has four-momentum $k_-$ before the interaction, while in diagram (B$_+$) it has four-momentum $k_+$ after the interaction, where $k_\pm=k\pm q$ with $k=\{E_k,{\bf k}\}$.  Therefore, in both cases particle 1 is off-shell unless $q=0$, so that, as we make the expansion (\ref{eq:quad}) needed to calculate the quadrupole moment, we probe the behavior of the vertex function when both particles are off-shell.   However, even if there were no interaction current, there would still be contributions of this type  from the vertex function itself.  It turns out that the interaction current cancels some of theses contributions, and  this subtracted vertex function is denoted by $\widehat \Gamma_{\rm BS}$. It depends on wave functions generically denoted by $\widehat z$.  I have made no attempt to separate the contributions of interaction current from that of the vertex function itself, so these contributions include both effects.  The contribution of these terms to the normalization (Table \ref{tab:Nnbrs}) gives a large contribution of almost 6\% (3\%) to the quadrupole moment from WJC-1 (WJC-2), and the additional contributions from $Q_{V_1}$ is about 10 times smaller.

The $Q_{\rm int}$ interference  term includes contributions from the product of the $w$ and $v_t$ wave functions not continued in the other terms.  Note that it makes a large contribution of almost $-1.5\%$ to the quadrupole moment for WJC-1, and a very small contribution for WJC-2.  This term is largely the cause of the small WJC-1 result.

 The $Q_{P}$ and $N_P$ terms include contributions from the square of the P-states and are quite small for both models.
 
 Finally, the interesting $Q_{\chi}$ term is the interference between the $v_s$ P-state and the combination of negative energy helicity states $z_\delta$.  It is quite small in both models, but for WJC-1 it is larger than the estimated theoretical error of 0.0001, and is therefore included.

Looking at the cumulative totals shown in Fig.~\ref{fig:two}, we conclude that the the result for model WJC-2 is quite close to the experimental value, and well given by the normalization correction, $N_{CST}$, alone.  The case for model WJC-1 is quite different however; here the additional corrections shown in Table \ref{tab:Qnbrs} reduce the quadrupole moment to an unacceptably low value, due largely to the single term $Q_{\rm int}$.  I discuss the significance of these results in the next section.

\section{conclusions and outlook} \label{sec:imp}

In this paper I present an approximate calculation (accurate to about 0.1\%) of the deuteron quadruple moment for two recent models that both give a high precision fit ($\chi^2/{\rm datum}\simeq 1$) to the 2007 $np$ data base below 350 MeV lab energy.   Model WJC-1, designed to give the best fit possible, has 27 parameters, $\chi^2/{\rm datum}\simeq 1.06$, and a large $\nu_{\sigma_0}=-15.2$.  Model WJC-2, designed to give a excellent fit with as few parameters as possible,  has only 15 parameters, $\chi^2/{\rm datum}\simeq 1.12$, and a smaller $\nu_{\sigma_0}=-2.6$.  Both models also predict the correct triton binding energy \cite{Gross:2008ps,Stadler:1996ut} and give a the same magnetic moment (with the uncertainty of 0.001) about 1\% larger than the experimental value.  

Until now, the major distinction between these two models has been their deuteron momentum distributions.  Model WJC-1 gives a much harder distribution than WJC-2 \cite{Ford:2014yua} and other models \cite{Ford:2014yua,Ehlers:2014jpa,Ethier:2014bua}, but since the momentum distribution is not an observable, it may be inappropriate  to use this as a means of distinguishing between them.  The prediction of the quadrupole moment  presented in this paper clearly  favors WJC-2.  The simplicity of model WJC-2, with only 15 parameters and a pure pseudo vector $\pi NN$ coupling, might also favor WJC-2, even though the $\chi^2$ of fit to the $np$ database is very slightly larger than that of WJC-1 (1.12 vs.~1.06).  Perhaps a calculation of the form factors, planned for the last paper in this series, will be definitive.  

How close we can expect the agreement to be between experimental data and the CST?  Perhaps agreement to about 1\% should be expected of the theory is to be taken seriously, and (in agreement with Machleidt \cite{Machleidt:2000ge}) I take the error of $-2.5\%$ in the WJC-1 prediction to be a serious problem.  On the other hand, should the error of  $-0.8\%$ in the WJC-2 prediction be accepted?  One answer is that the $\chi$EFT prediction is comparable, and this claims to be a theory and not just a model.  If we want an exact prediction, recall that the deuteron binding energy and the $^1S_0$  scattering lengths were already constrained when fitting the $np$ data base \cite{Gross:2008ps}, so perhaps the deuteron quadrupole moment could also be constrained at the same time.  Since model WJC-2 agrees so closely without this constraint, perhaps it could be included without seriously degrading the $\chi^2$.  These possibilities await future study.

\acknowledgements
It is a pleasure to thank Rocco Schiavilla for stimulating conversations about the $\chi$EFT predictions.  This material is based upon work supported by the by Jefferson Science
Associates, LLC, under U.S. DOE Contract No. DE-AC05-06OR23177.
 
\appendix*

\section{DETAILS OF THE CALCULATION}

For any quantity not defined in the discussion below, refer to Ref.~II
 
 \subsection{Diagrams (A) and (A$_\pm$)}

\subsubsection{Exact expressions}

The quadrupole form factor, $G_Q$, is obtained directly from difference between ${\cal J}_1$ and ${\cal J}_2$.  Using Eq.~(\ref{eq:dffmatrix}) and the results from Ref.\ II, this  is
\begin{widetext}
\bea
4D_0\eta\,G_Q(Q^2)\Big|_{{\rm A}+{\rm A}_\pm}&& =e_0F_1(Q^2)\int_k\Big\{f_0(p_+,p_-)\,\delta {\cal A}_1(k,Q)-\frac{h_+}{h_-}\,\delta{\cal A}^{(2)}_1(k,Q)-\frac{h_-}{h_+}\,\delta {\cal A}_1^{(2)}(k,-Q)\Big\}
\nonumber\\&&
+e_0 F_2(Q^2)\int_k \Big\{ f_0(p_+,p_-)\, \delta {\cal A}_2(k,Q) -\frac{h_+}{h_-}\,\delta{\cal A}^{(2)}_2(k,Q)+\frac{h_-}{h_+}\,\delta {\cal A}^{(2)}_2(k,-Q)\Big\}\nonumber\\&&
+e_0 F_3(Q^2)\int_k \frac{g_0(p_+,p_-)}{4m^2} \,\delta {\cal A}_3(k,Q)\, .
\label{eq:GQfromA}
\eea
%
where $\delta\,{\cal A}_i$ are differences of the traces ${\cal A}_{n,i}$ defined in Ref.~II
\bea
\delta {\cal A}_i(k,Q)&\equiv&{\cal A}_{1,i}(\Psi_+,\Psi_-)-{\cal A}_{2,i}(\Psi_+,\Psi_-)
\nonumber\\
\delta {\cal A}^{(2)}_{i}(k,Q)&\equiv&{\cal A}_{1,i} (\Psi_+,\Psi^{(2)}_-)-{\cal A}_{2,i} (\Psi_+,\Psi^{(2)}_-) .
\label{eq:Qdiffs}
\eea
%
Introducing the convenient averages
\bea
\overline{A}_{i}(k)&=&\lim_{Q^2\to0}\frac{m_d}{Q^2}\,\frac12\int_{-1}^1dz\;\delta{\cal A}_i(k,Q)
\nonumber\\
\overline{A}^{(2)}_{i\pm}(k)&=&\lim_{Q^2\to0}\frac{m_d}{Q^2}\,\frac12\int_{-1}^1dz\;\delta{\cal A}^{(2)}_i(k,\pm Q)\, , \label{eq:QlimA}
\eea
the contributions of diagrams A and A$_\pm$ to the quadrupole moment is written
\bea
Q_A
=&&e_0\int \frac{k^2dk}{2\pi^2} \frac{m}{E_k}\Big\{f_{00}\,\overline{A}_{1}(k) -\overline{A}^{(2)}_{1+}(k) -\overline{A}^{(2)}_{1-}(k) \Big\}
\nonumber\\&&
+e_0\,\kappa_s\int \frac{k^2dk}{2\pi^2} \frac{m}{E_k} \Big\{f_{00}\,\overline{A}_{2}(k) -\overline{A}^{(2)}_{2+}(k) -\overline{A}^{(2)}_{2-}(k)\Big\}
+e_0\int \frac{k^2dk}{2\pi^2} \frac{m}{E_k} \frac{g_{00}}{4m^2} \overline{A}_{3}(k) \, ,  \label{eq:quad1}
\eea
where $f_{00}$ and $g_{00}$ are coefficients of the off-shell nucleon current defined in Eq.~(3.24) of Ref.~II.

To work out the limits (\ref{eq:QlimA}), expand the differences (\ref{eq:Qdiffs}) to order $Q^2$.  Making the approximation $m_d\simeq 2m$  gives
\bea
\delta {\cal A}_1(k,Q)&=&\frac{2k^2}{m^3} P_2(z)\bigg\{B_+B_-E_k-4C_+C_-(2m-E_k)+4D_+D_-E_k+2(A_+D_-+D_+A_-)m
\nonumber\\&&-2(B_+C_-+C_+B_-)(E_k-m)-2(B_+D_-+D_+B_-)m\bigg\}
\nonumber\\&&
+\frac{k_zQ}{2m^3}\bigg\{\big[A_+(B_--2C_-)-(B_+-2C_+)A_-\big]m+2(A_+D_--D_+A_-)E_k
\nonumber\\&&
-2(B_+C_--C_+B_-)(E_k-m)-4(C_+D_--D_+C_-)(E_k-m)\bigg\}
+\frac{Q^2}{2m^3}E_k\Big\{C^2\bigg(1-\frac{2k^2}{3m^2}\Big)
\nonumber\\&&
\qquad-\Big(\frac14B^2+D^2\Big)\Big(1+\frac{2k^2}{3m^2}\Big)+\frac12AB-AC+BC\frac{2k^2}{3m^2}+2CD\Big(1-\frac{m}{E_k}-\frac{2k^2}{3E_km}\Big)\bigg\}
\nonumber\\
\delta {\cal A}_2(k,Q)&=&\frac{k_zQ}{m^3}\bigg\{(A_+B_--B_+A_-)E_k-2(A_+C_--C_+A_-)(E_k-2m)-2(A_+D_--D_+A_-)m
\nonumber\\&&
+2(B_+C_--C_+B_-)(E_k-m)-2(B_+D_--D_+B_-)\frac{k^2}{m}P_2(z)
\nonumber\\&&-4(C_+D_--D_+C_-)\Big[E_k-m-\frac{k^2}{m}P_2(z)\Big]\bigg\}
-\frac{Q^2}{2m^2}\bigg\{A^2-(4C^2-2BC-4CD)\Big(1-\frac{E_k}{m}+\frac{2k^2}{3m^2}\Big)
\nonumber\\&&
-AB\Big(1+\frac{2k^2}{3m^2}\Big)-2AC\Big(1-\frac{2k^2}{3m^2}\Big)+2AD\frac{E_k}{m}\bigg\}
\label{eq:deltaA}
\eea
where $P_2(z)$ is the $\ell=2$ Legendre polynomial with $z=k_z/k$ the cosine of the polar angle, and $Z_\pm=Z(R_\pm)$ (where $Z$ is a generic name for the $A,B,C$, or $D$ invariants defined in Ref.~II) and $R_\pm$ is the covariant generalization of the magnitude of rest frame three-momentum $|{\bf k}|$ for the outgoing ($R_+$) and incoming ($R_-$) deuteron states.  From Ref.~II, these arguments, expanded to order $Q^2$, are 
\bea
R_{\pm}&&\simeq\Big[{\bf k}^2\mp \frac{k_zQE_k}{m_d} +\eta\left(E_k^2+k_z^2\right)\Big]^{1/2}
\simeq k\mp z\,Q\,\frac{E_k}{2m_d} +\frac{\eta}{2k} (E_k^2- m^2z^2)
\eea
where now $|{\bf k}|\to k$.  In calculating the average $\overline A_1(k)$, the first term  will get contributions of order $Q^2$ from the expansions of the wave functions, but only terms proportional to $z^2 P_2(z)$ will survive.  Hence, for arbitrary $\{X,Y\}=\{A,B,C,D\}$, the expansion needed is
\bea
X_+Y_-&=&X(R_+)Y(R_-)\to -z^2 \eta\Big\{(X'Y+XY')\frac{m^2}{2k}+X'Y'E_k^2-\frac12(X''Y+XY'')E_k^2\Big\}
\quad
\eea  
where 
 $X=X(k)$, $X'=dX(k)/dk$, etc.   
Only derivative terms contribute to the terms proportional to $k_xQ$, and for these we need
\bea
X_+Y_-\to -\frac{zQE_k}{2m_d}(X'Y-XY')\, .
\eea
Making these substitutions and continuing to let $m_d\to 2m$, reduces the averages  (\ref{eq:QlimA}) to
\bea
\overline{A}_{1}(k)&=&-\frac{k^2E^2_k}{30m^4}\Big\{(B'^2+4D'^2-B''B-4D''D)E_k -2(2B'C'-B''C-C''B)(E_k-m) +4(C'^2-C''C)(E_k-2m)
\nonumber\\&&\qquad
+2(2A'D'-A''D-D''A+B''D+D''B-2B'D')m\Big\}
\nonumber\\&&
+\frac{k}{30m^3}\Big\{\big[5(AB'-A'B)-B'B+10(A'C-AC')-4D'D\big]mE_k+2(B'D+D'B)m^2
-4C'C(E_k-2m)m
\nonumber\\&&\qquad
+20(C'D-D'C)E_k(E_k-m)-2C'B(E_k-m)(5E_k-m)+2B'C(E_k-m)(5E_k+m)
\nonumber\\&&\qquad
-2A'D(5E_k^2+m^2)+2D'A(5E_k^2-m^2)\Big\}
\nonumber\\&&
-\frac1{12m^4}\Big\{(B^2+4D^2)(2E_k^2+m^2)E_k-4BDm(2E_k^2+m^2)-6(AB-2AC)E_km^2-8BCk^2E_k
\nonumber\\&&\qquad
+8CDm(E_k-m)(2E_k-m)+4C^2E_k(2E_k^2-5m^2)\Big\} \label{eq:A1}
\eea
\bea
\overline{A}_{2}(k)&=&\frac{4k^3 E_k}{15m^4}\Big\{B'D-D'B-2C'D+2D'C\Big\}-\frac{kE_k}{3m^3}\Big\{(A'B-B'A)E_k-2(A'D-D'A)m
\nonumber\\&&\qquad
+2(B'C-C'B-2C'D+2D'C)(E_k-m) -2(A'C-C'A)(E_k-2m)\Big\}
-\frac1{3m^3}\Big\{3A^2m^2+6ADE_km
\nonumber\\&&\qquad
-AB(2E_k^2+m^2) +2AC(2E_k^2-5m^2) +2(BC-2C^2+2CD)(E_k-m)(2E_k-m)\Big\}
\label{eq:A2}
\eea
\end{widetext}
where $\overline A_{3}(k)$ is $\overline A_{1}(k)$ with $A\to F$, etc.  
As expected, $\overline{A}_{2}(k)$ includes no terms involving $Z''$ or $Z'^2$ because it is already ${\cal O}(Q)$ without expansions of the wave functions. 

\subsubsection{Leading terms in momentum space}

Equations (\ref{eq:A1}) and (\ref{eq:A2}) give the exact results for the quadrupole moment, and can be easily evaluated numerically.  However, our goal here is to obtain some insight into the physical content of the result, and to this end it is sufficient to  compute the quadrupole moment to an accuracy of about $0.1\%$ as we did for the magnetic moment in Ref.~II. This is done by expanding the exact results in terms of the four deuteron wave functions $z_\ell= \{u,w,v_t,v_s\}$ (where  $z_\ell$ is the generic name for any of the wave functions, and the expansions were  given in Ref.~II), and   retaining only the leading terms, as defined in Ref.~II.  These leading terms are obtained by expanding the coefficients of the leading  products of the wave functions $u$ and $w$ to order $k^2/m^2$, and expanding coefficients of all products involving P-state wave functions to order $k/m$.  In comparing derivative terms, $z_\ell'', z_\ell'/k,$ and $z_\ell/k^2$ are considered to be of the same order.   Pulling out an overall factor of $E_k$, and integrating by parts to remove all of the double derivatives and to make other simplifications, gives
\begin{widetext}
\bea
\overline{A}_{1}(k)&\simeq&  2 \pi^2 \frac{E_k}{m} \frac{m^2}{10}\bigg\{-4\sqrt{2}\,u'w'-2w'^2+2v_t'^2-4v_s'^2-12\sqrt{2}\,\frac{u' w}{k}
-\frac1{k^2}\Big[12w^2-4 v_t^2+8v_s^2\Big] 
+\frac{k}{\sqrt{3}\,m}A_{\rm int} + \Delta A_1\bigg\}
\nonumber\\
\overline{A}_{2}(k)&\simeq& 2 \pi^2 \frac{E_k}{m}\frac{m^2}{10}\bigg\{ \frac{2k}{\sqrt{3}\,m}A_{\rm int}+\Delta A_2\bigg\}
\nonumber\\
\overline{A}_{3}(k)&\simeq& 2 \pi^2 \frac{E_k}{m} \frac{4m^4}{10}\bigg\{2\Big[v_t'^2+\frac{2v_t^2}{k^2}\Big]-4\Big[v_s'^2+\frac{2v_s^2}{k^2}\Big]\bigg\}
\label{eq:A12lead}
\eea
\end{widetext}
where the interference terms, multiplied by a factor of $k/m$, are
\bea
A_{\rm int}&=&-\frac2{k}\Big\{5u'(\sqrt{2}v_t+2v_s)-4w\Big(v_t'-\frac{v_t}{k}\Big)
\nonumber\\&&
-\sqrt{2}w\Big(v_s'-\frac{v_s}{k}\Big)\Big\} \label{eq:Aint}
\eea
and $\Delta A$ is the $k^2/m^2$ correction to the leading terms
\bea
\Delta A_1&=&\frac{k^2}{m^2}\Big[4\sqrt{2} \, u'w'+2w'^2+\frac{6\sqrt{2}\,u'w}{k}+\frac{3w^2}{k^2}\Big]\qquad\quad
\nonumber\\
\Delta A_2&=&\frac{6 k}{m^2}\Big[ 2\sqrt{2} u' w +\frac1{k}(\sqrt{2}\,uw+w^2)\Big].
\eea
Note that the interference terms, smaller by one power of $k/m$, might be ignored, and, as it turns out, the contributions from the (B) diagrams contribute interference terms that larger by a factor of $m^2/k^2$.  The contributions from $A_{\rm int}$ will therefore be ignored when the contributions  (A) and (B) diagrams are combined.

As discussed in Ref.~II, the A$_\pm$ contributions to the interaction currents are obtained by substituting $\Psi^{(2)}$ wave functions for $\Psi$ wave functions in the initial state, giving rise to the $A^{(2)}$ terms in Eq.~(\ref{eq:GQfromA}).   These  can be constructed directly from  (\ref{eq:A12lead}).  Since the  $\Psi^{(2)}$ contributions are already small, they will be kept only to leading order, so that any contributions that might have come from  $A_{\rm int}$ will be discarded.  The $\delta A_2$ terms in  Eq.~(\ref{eq:A12lead}) can therefore be ignored.  To find the  $\Psi^{(2)}$ contributions $\delta A_1^{(2)}$, recall from Ref.~II that the helicity traces ${\cal A}_{n,1}$ from which  $\delta A_1$ is calculated satisfy the symmetry relation (for $n=1,2$)
\bea
{\cal A}_{n.1}(\Psi_1\Psi_2)={\cal A}_{n.1}(\Psi_2\Psi_1)\Big|_{q\to-q}\, . \label{eq:Asymm}
\eea
Note that a typical term in the expansion  (\ref{eq:deltaA}) satisfies this symmetry, and is of the form
\bea
\left<X_+Y_-\right>&\to&( P_2(z)c_0+Q^2c_2)(X_+Y_-+Y_+X_-)
\nonumber\\&&
+k_zQc_1(X_+Y_--Y_+X_-)
\eea
where the $c_i$ include all of the additional factors present in the expansions.  Replacing the initial state by $\Psi^{(2)}$, and exploiting this symmetry, means that the typical $XY$ contribution to the $\delta A_1^{(2)}(k,Q)$ term in (\ref{eq:GQfromA})  becomes   
\bea
\left<X_+Y_-^{(2)}\right>\Big|_Q&\to& (P_2(z)c_0+Q^2c_2)(X_+Y_-^{(2)}+Y_+X_-^{(2)})
\nonumber\\&&
+k_zQC_1(X_+Y_-^{(2)}-Y_+X_-^{(2)}).
\eea
Adding the second contribution in Eq.~(\ref{eq:GQfromA}), $\delta A_1^{(2)}(k,-Q)$, gives a combined result
\begin{widetext}
\bea
\left<X_+Y_-^{(2)}\right>\Big|_Q+\left<X_+Y_-^{(2)}\right>\Big|_{-Q}&\to& (P_2(z)c_0+Q^2c_2)(X_+Y_-^{(2)}+X_+^{(2)}Y_-+Y_+X_-^{(2)}+Y_+^{(2)}X_-)
\nonumber\\&&
+k_zQC_1(X_+Y_-^{(2)}+X_+^{(2)}Y_--Y_+X_-^{(2)}-Y_+^{(2)}X_-)
\eea
showing that all terms are obtained by the expected substitution $XY\to XY^{(2)}+X^{(2)}Y$ where either $X$ or $Y$ may contain one or two derivatives.  The contributions from $\Psi^{(2)}$ therefore reduce to
\bea
\overline{A}_{1+}^{(2)}(k)+\overline{A}_{1-}^{(2)}(k)&\simeq&  2 \pi^2 \frac{E_k}{m} \frac{m^2}{10}\bigg\{-4\sqrt{2}\,(u'{w^{(2)}}'+{u^{(2)}}'w')-4w'{w^{(2)}}'+4v_t'{v_t^{(2)}}'-8v_s'{v_s^{(2)}}'
\nonumber\\&&
-\frac{12\sqrt{2}}{k}({u^{(2)}}'w +u' w^{(2)})
-\frac1{k^2}\Big[24ww^{(2)}-8 v_tv_t^{(2)}+16v_sv_s^{(2)}\Big] 
\bigg\}\, . \qquad \label{eq:QAaterms}
\eea

Finally, the contributions from the derivatives of the strong from factor, $h$, expressed in terms of $a(p)$ defined in Sec.~\ref{sec:cal} (and Eq.~(3.25) of Ref.~II), are extracted from contributions from $f_{00}$ and $g_{00}$.  
These terms will be simplified by integrating by parts as we did for the leading contributions (\ref{eq:A12lead}).  In doing this integration, we use the fact that $a(p^2)$ is a function of $p^2=m^2-m_d(2E_k-m_d)\simeq m^2-2k^2$, so that  $da(p^2)/(dk)$ is suppressed by one power of $k$ and can be ignored.  The contributions from $\overline{A}_2$ are not of leading order, so that the $a(p^2)$ contributions that might have come from this term can be neglected.  The leading contributions from  $\overline{A}_1$ and $\overline{A}_3$ combine to give
\bea
Q_A\Big|_{h'}&=&e_0\int_0^\infty\frac{k^2 dk}{2\pi^2}\frac{m}{E_k}\,2a(p^2)\Big\{m_d(2E_k-m_d)\overline{A}_1(k)-\overline{A}_3(k)\Big\}
\nonumber\\
&\simeq&-e_0\frac{m^2}{10}\int_0^\infty k^2 dk\,2a(p^2)\bigg\{2k^2\Big[4\sqrt{2}\,u'w'+2w'^2+\frac{12\sqrt{2}u'w}{k}+\frac{6w^2}{k^2}\Big]+8m^2\Big[v_t'^2+\frac{2v_t^2}{k^2}-2v_s'^2-\frac{4v_s^2}{k^2}\Big]\bigg\},\qquad\quad
\label{eq:Qhprime}
\eea
\end{widetext}
where, when integrating the $u,w$ terms by parts, use the fact that the volume element is $k^4dk$ (instead of $k^2dk$ as it was for $\overline{A}_1$), giving integrated contributions to (\ref{eq:Qhprime}) that differ from those shown in (\ref{eq:A12lead}).

\subsubsection{Leading terms in coordinate space}

In view of the rich history and importance of this quantity, it is instructive to  cast the leading contributions into coordinate space where they have a simple and familiar form.   

To aid transforming the terms of ${\cal O}(1)$, use the general identities (for arbitrary $\ell$ and $\ell'$)
\begin{widetext}
\bea
&&a\int_0^\infty dk \frac{d}{dk}\big(k \,z_\ell z_{\ell'}\big)=a\int_0^\infty k^2dk\Big(\frac{z_\ell' z_{\ell'}+z_\ell z_{\ell'}'}{k}+\frac{z_\ell z_{\ell'}}{k^2}\Big)=0
\nonumber\\
&&b\int_0^\infty dk \frac{d^2}{dk^2}\big(k^2 z_\ell z_{\ell'}\big)=b\int_0^\infty k^2dk\Big(z_\ell'' z_{\ell'}+z_\ell z_{\ell'}''+2z_\ell' z_{\ell'}'+\frac{4(z_\ell' z_{\ell'}+z_\ell z_{\ell'}')}{k}+\frac{2z_\ell z_{\ell'}}{k^2}\Big)=0
\nonumber\\
&&c_\ell\int_0^\infty dk \frac{d}{dk} (k^2z_\ell'z_{\ell'})=c_\ell\int_0^\infty k^2dk \big(z_\ell'' z_{\ell'}+z_\ell' z_{\ell'}'+\frac{2z_\ell'z_{\ell'}}{k}\big)=0
\, . \label{eq:zlidentities}
\eea
Using these in the calculation of the $uw$ terms  gives
\bea
Q_A\big|_{u,w}&=&e_0\int_0^\infty\frac{k^2\,dk}{2\pi^2}\frac{m}{E_k}\overline  A_1(k)\big|_{u,w}=\frac{e_0\,m^2}{5\sqrt{2}}\int_0^\infty k^2\,dk\Big\{-4u'w' -12\frac{u'w}{k}\Big\}
\nonumber\\
&=&\frac{e_0\,m^2}{5\sqrt{2}}\int_0^\infty k^2\,dk\Big\{(b+c_0)u''w +(b+c_2)u w'' +(2b+c_0+c_2-4)u'w'
\nonumber\\&&\qquad\qquad 
+(2c_0+4b+a-12)\frac{u'w}{k}
+(2c_2+4b+a)\frac{uw'}{k}+(2b+a)\frac{uw}{k^2}\Big\}
\nonumber\\
&=&e_0\frac{4m^2}{5\sqrt{2}}\int_0^\infty k^2\,dk\Big(u''-\frac{u'}{k}\Big)w
\label{eq:Qlead}
\, ,
\eea
\end{widetext}
where, for any $c_0$, $a=2c_0-8,b=4-c_0, c_2=c_0-4$.  To reduce this further, use the fact that the momentum and position space wave functions are related by the spherical Bessel transforms
\bea
z_\ell(k)&=&\sqrt{\sfrac2{\pi}}\int_0^\infty r d{r}\,j_\ell(kr)\,z_\ell(r)
\nonumber\\
\frac{z_\ell(r)}{r}&=&\sqrt{\sfrac2{\pi}}\int_0^\infty k^2dk\,j_\ell(kr)\,z_\ell(k)
\label{eq:besseltrans}
\eea
where $j_\ell$ is the spherical Bessel function of order $\ell$, satisfying the equation
\bea
\Big(\frac{d^2}{dx^2}+\frac2{x}\frac{d}{dx}-\frac{\ell(\ell+1)}{x^2}+1\Big)j_\ell(x)=0
\label{eq:besseleq}
\eea
with the convenient recursion relations
\bea
j_\ell(z)=z^\ell\left(-\frac{1}{z}\frac{d}{dz}\right)^\ell\frac{\sin z}{z}. \label{eq:recursion}
\eea
and the normalization condition
\bea
\int_0^\infty k^2dk j_\ell(kr)j_\ell(kr')=\frac{\pi}{2r^2}\delta(r-r')\, . \label{eq:besselnorm}
\eea
Hence, the Bessel transform (\ref{eq:besseltrans}), and the recursion relation (\ref{eq:recursion}), give 
\bea
u''(k)-\frac{u'(k)}{k}&=&\sqrt{\frac2\pi}\int_0^\infty rdr\Big(\frac{d^2}{dk^2}- \frac1{k}\frac{d}{dk}\Big)\,j_0(kr)u(r)
\nonumber\\
&=& \sqrt{\frac2\pi}\int_0^\infty r^3dr\,j_2(kr)u(r)
\eea
reducing (\ref{eq:Qlead}) to
\bea
Q_A\big|_{u,w}&=&e_0\frac{4m^2}{5\sqrt{2}}\left[\frac{2}{\pi}\right]\int_0^\infty k^2\,dk\int_0^\infty r^3 dr j_2(kr) u(r)
\nonumber\\
&&\times \int_0^\infty r'dr'j_2(kr')w(r')
\nonumber\\
&=&e_0\frac{4m^2}{5\sqrt{2}}\int_0^\infty r^2 dr \,u(r)w(r)
\label{eq:Auwterms}
\eea

\begin{widetext}
The leading $w^2$ term can be similarly reduced.  Using the identities (\ref{eq:zlidentities}) and (\ref{eq:besseleq}) gives
\bea
Q_A\big|_{w^2}&=&-e_0\frac{m^2}{5}\int_0^\infty k^2\,dk\Big\{w'^2+\frac{6w^2}{k^2}\Big\}
\nonumber\\
&=&-e_0\frac{m^2}{5}\int_0^\infty k^2\,dk\Big\{(2b+c_2)ww''+(2b+c_2+1)w'^2+(2c_2+8b+2a)\frac{ww'}{k}+(2b+a+6)\frac{w^2}{k^2}\Big\}
\nonumber\\
&=&e_0\frac{m^2}{5}\int_0^\infty k^2\,dk\Big(w''+\frac{2w'}{k}-\frac{6w}{k^2}\Big)w=-e_0\frac{m^2}{5}\int_0^\infty r^2 dr \,w^2(r)
\label{eq:Aw2terms}
\eea
where $c_2=-1-2b$ and $a=-2b$.  Similarly, using  (\ref{eq:besseleq}) for $\ell=1$ the leading P-wave terms become
\bea
Q_A\big|_{P^2}&=&e_0\frac{m^2}{5}\int_0^\infty k^2\,dk\Big\{v_t'^2-2v_s'^2+\frac{2v_t^2-4v_s^2}{k^2}\Big\}
\nonumber\\
&=&e_0\frac{m^2}{5}\int_0^\infty k^2\,dk\Big\{c_1v_tv_t'' +(c_1+1)v_t'^2 +2c_1\frac{v_tv_t'}{k}+\frac{2v_t^2}{k}
+ c_1'v_sv_s''+(c_1'-2)v_s'^2+2c_1'\frac{v_sv_s'}{k}-\frac{4v_s^2}{k}\Big\}
\nonumber\\
&=&e_0\frac{m^2}{5}\int_0^\infty k^2\,dk\Big\{-\Big(v_t''+\frac{2v_t'}{k}-\frac{2v_t^2}{k^2}\Big)v_t+2\Big(v_s''+\frac{2v_s'}{k}-\frac{2v_s}{k^2}\Big)v_s\Big\}=e_0\frac{m^2}{5}\int_0^\infty r^2 dr\,(v_t^2-2v_s^2),\qquad
\label{eq:AP2terms}
\eea
where $c_1=-1$ and $c_1'=2$.  

Summing (\ref{eq:Auwterms}), (\ref{eq:Aw2terms}), and (\ref{eq:AP2terms}) gives the leading contribution to to the quadrupole moment from the (A) diagrams.   
Multiplying this dimensionless quantity by $1/m_d^2\simeq1/(4m^2)$  gives the physical quadrupole moment for a deuteron with unit charge, 
\bea
Q_d\Big|_0=\frac1{4m^2}Q_A\Big|_0= e_0\frac{\sqrt{2}}{10}\int_0^\infty r^2dr\Big\{u(r)w(r)-\frac1{\sqrt{8}}\big[w^2(r)-v_t^2(r)+2v_s^2(r)\big]\Big\}\, .
\eea
Since $e_0=1/2$, this is {\it one-half\/} of the RIA result, and it agrees with the leading terms in Eq.\ (1.16) of Ref.\ \cite{Arnold:1979cg}; note that the $uw$ and $w^2$ terms are identical to 1/2 of the familiar non-relativistic result (the other 1/2 comes from the B diagrams).

Next we evaluate the terms of order $k/m$, which arise only from interference between the leading S and D-state components and the smaller P-state components.  These are the $A_{\rm int}$  terms defined in Eq.\ (\ref{eq:Aint}).   Their contribution is
\bea
Q_A\Big|_1&=&e_0(1+2\kappa_s)\frac{m}{10\sqrt{3}}\int_0^\infty k^2\,dk\Big\{-10\sqrt{2}\,u'v_t-20\,u'v_s+8\Big[wv_t'-\frac{wv_t}{k}\Big]
+2\sqrt{2}\Big[wv_s'-\frac{wv_s}{k}\Big]\Bigg\}\, .\qquad\qquad
\eea
Next, using
\bea
u'(k)&=&\sqrt{\frac2\pi}\int_0^\infty r^2dr\Big( \frac1{r}\frac{d}{dk}\Big)\,j_0(kr)u(r)= -\sqrt{\frac2\pi}\int_0^\infty r^2dr\,j_1(kr)u(r)
\nonumber\\
v'(k)  -\frac{v(k)}{k}&=&\sqrt{\frac2\pi}\int_0^\infty r^2dr\Big(\frac1{r}\frac{d}{dk}-\frac1{kr}\Big)\,j_1(kr)v(r)= -\sqrt{\frac2\pi}\int_0^\infty r^2dr\,j_2(kr)v(r) \label{eq:H43}
\eea
and the normalization condition (\ref{eq:besselnorm}), these terms give the following contributions to the quadrupole moment
\bea
Q_d\Big|_1=\frac1{4m^2} Q_A\Big|_1=e_0(1+2\kappa_s)\frac{1}{2\sqrt{3}}\int_0^\infty dr\frac{r}{m}\Bigg\{u(r)\Big[\frac1{\sqrt{2}}v_t(r)+ v_s(r)\Big] -\frac25 w(r)\Big[v_t(r)+\frac1{2\sqrt{2}}\,v_s(t)\Big]\Bigg\}\, .
\eea
Multiplying by 2 gives the RIA result, which also agrees with Ref.\ \cite{Arnold:1979cg}
.

\subsection{Diagrams (B) and (B$_\pm$)}

Using  results from Ref.\ II, the contributions from diagrams B plus B$_\pm$ to the quadrupole moment are
\bea
4D_0\eta\,G_Q(Q^2)\Big|_{{\rm B}+{\rm B}_\pm}\!\!=e_0 F_1(Q^2)\int_k&&\Bigg\{\frac{m}{\kappa_z} \left[\frac{\delta{\cal B}_1(k_0,Q)}{k_0}\Big|_{-}- \frac{\delta{\cal B}_1(k_0,Q)}{k_0}\Big|_{+}\right] -
\frac1{m} \Big[\delta{\cal C}_1(k,Q)+\delta{\cal C}_1(k,-Q)\Big]\Bigg\}
\nonumber\\
+e_0 F_2(Q^2)\int_k&&\Bigg\{ \frac{m}{\kappa_z}\left[\frac{\delta{\cal B}_2(k_0,Q)}{k_0}\Big|_{-}
- \frac{\delta{\cal B}_1(k_0,Q)}{k_0}\Big|_{+}\right]  - 
\frac1{m} \Big[\delta{\cal C}_2(k,Q)-\delta{\cal C}_2(k,-Q)\Big]\Bigg\},\qquad\quad
\label{eq:QB&V1}
\eea
\end{widetext}
where $|_\pm\to |_{k_0=E_\pm}$ with $E_\pm=\sqrt{m^2+({\bf k}\pm {\bf q}/2)^2}$, $\kappa_z\equiv {\bf k}\cdot{\bf q}/E_k=k_zQ/E_k$, and the $\delta{\cal B}_i$ and $\delta{\cal C}_i$ differences are
\bea
\delta{\cal B}_i(k_0,Q)&=&{\cal B}_{1,i}(k_0)-{\cal B}_{2,i}(k_0)
\nonumber\\
\delta{\cal C}_i(k,Q)&=&{\cal C}_{1,i}(\Gamma\Gamma_{\rm off})-{\cal C}_{2,i}(\Gamma\Gamma_{\rm off})
\label{eq:Bdiffs}
\eea
where the traces ${\cal B}_{n,i}$ and ${\cal C}_{n,i}$ were defined in Ref.~II.  
Introducing the averages
\bea
\overline{{\cal B}}_{i}(k_0)&=&\lim_{Q^2\to0}\frac{m_d}{Q^2}\frac12\int_{-1}^1dz\;\frac{E_k}{k_0}\,\delta{\cal B}_i(k_0,Q)
\nonumber\\
\overline{{\cal C}}_{i\pm}(k)&=&\lim_{Q^2\to0}\frac{m_d}{Q^2}\frac12\int_{-1}^1dz\;\delta{\cal C}_i(k,\pm Q)\, , \label{eq:QlimB}
\eea
and the combinations
\bea
\overline{B}_i(k)&\equiv& \frac{m}{k_zQ}\Big(\overline{\cal B}_i\big|_-
-\overline{\cal B}_i\big|_+\Big)
\nonumber\\
\overline{C}_i(k)&\equiv&\frac1{m}\big(\overline{\cal C}_{i+}+\overline{\cal C}_{i-}\Big)
\eea
the contributions of diagrams B plus B$_\pm$ to the quadrupole moment become
\bea
Q_B&=&e_0\int\frac{k^2dk}{2\pi^2}\frac{m}{E_k}\Big\{
\overline{B}_1(k)+\kappa_s\overline{B}_2(k)
\nonumber\\&&\qquad\qquad\qquad
-\overline{C}_{1}(k)-\kappa_s\overline{C}_{2}(k)\Big\} .
\label{eq:quadB}
\eea
We will refer to the ${\cal B}$ contributions as the ``singular'' terms, even thought the singularity at $k_z=0$ is cancelled by the subtraction of two terms evaluated at $k_0=E_\pm$.  The ${\cal C}$  contributions are individually finite and depend on the vertex function with {\it both\/} nucleons off shell, $\Gamma_{\rm off}$, introduced in Eq.~(2.12) of Ref.~II.  

\subsubsection{Evaluation of the singular terms}

At small $Q$, the factor $\overline{{\cal B}}_i/k_0$ can be expanded in a power series in  $(k_0-E_k)^n$, and the differences $\overline{{\cal B}}_i|_- - \overline{\cal B}_i|_+$   evaluated.   These differences, weighted by the factor $\kappa_z$,  cannot contribute to  the quadrupole moment if they are of higher order than $Q^2$.     Introducing 
\bea
E_\pm-E_k\simeq &\pm&\frac{k_z Q}{2E_k} +\frac{Q^2}{8E_k^3}(E_k^2-k_z^2)\mp\frac{k_zQ^3}{16E_k^5}(E_k^2-k_z^2)
\nonumber\\
&\equiv&  \epsilon_\pm  \, , \label{eq:epm}
\eea
these differences, up to order $Q^2$, are
\bea
&&\frac1{\kappa_z}(\epsilon_+  - \epsilon_-)\simeq \Big[1-\frac{Q^2}{8E_k^4}(E_k^2-k_z^2)\Big]\to \Big[1+\frac{z^2Q^2k^2}{8E_k^4}\Big]
\nonumber\\
&&\frac1{\kappa_z}(\epsilon_+^2  - \epsilon_-^2)\simeq \frac{Q^2}{4E_k^3}(E_k^2-k_z^2) \to-\frac{z^2Q^2k^2}{4E_k^3} 
\nonumber\\
&&\frac1{\kappa_z}(\epsilon_+^3  - \epsilon_-^3)\simeq \frac{k^2_zQ^2}{4E_k^2}
\to \frac{z^2Q^2k^2}{4E_k^2} , \label{eq:k0factor}
\eea
where contributions from all other powers of $(k_0-E_k)^n$ are negligible, and at order $Q^2$, only the $z^2Q^2$ terms will contribute, and explained below.  Expanding the coefficients of $\delta {\cal B}_i/k_0$ in a power series in $Q$, Eq.~(\ref{eq:k0factor}) shows that only the lowest order contribution from the term linear in $k_0-E_k$ can contribute to the terms of order $Q$ and $Q^2$, but that all three powers could, in principle contribute to the  term of order $Q^0$.  However, it turns out that the zeroth order term is accompanied by the Legendre polynomial $P_2(z)$, so that only the contributions proportional to $z^2$ will survive the integration over $z$ weighted by $P_2(z)$.
Recalling the definition of the reduced invariants  $X_+=h\widetilde X(\tilde R_+, R_0^+)$ and $Y_-=h\widetilde Y(\tilde R_-, R_0^-)$ (with $X,Y$ generic names for $F,G,H$, or $I$), with $h=h(\tilde p)$ the strong form factor (which for these contributions is a function of $\tilde p^2=(D_0-k_0)^2-k^2$), the contribution from a typical product of invariants $X_+Y_-$  has the form
\begin{widetext}
\bea
\frac{\delta{\cal B}_i}{k_0}\Big|_{XY}&=&P_2(z)\Big[B_{00,i}^{XY}+(k_0-E_k)B_{01,i}^{XY}\Big]h^2\widetilde X_+\widetilde Y_-  +P_2(z)\Big[(k_0-E_k)^2 B_{02,i}^{XY}+(k_0-E_k)^3B_{03,i}^{XY}\Big] X Y 
\nonumber\\
&&
+k_zQ\Big[B_{10,i}^{XY}(z^2)+(k_0-E_k)B_{11,i}^{XY}(z^2)\Big]h^2\widetilde X_+\widetilde Y_-
+ Q^2\Big[B_{20,i}^{XY}(z^2)+(k_0-E_k)B_{21,i}^{XY}(z^2)\Big]h^2\widetilde X_+\widetilde Y_- \qquad\label{eq:dBk0}
\eea
\end{widetext}
where the coefficient $B_{nm,i}^{XY}$ multiplies $Q^n(k_0-E_k)^m$.  All of these coefficients are independent of $Q$ and $k_0$, but may be a linear function of $z^2$, as indicated.  Note the factor of $P_2(z)$ multiplying the terms of ${\cal O}(Q^0)$, and that the form of the terms proportional to $(k_0-E_k)^{2,3}$ anticipates that the arguments of the invariants must be evaluated at $Q=0$;  the differences (\ref{eq:k0factor}) ensure that higher order terms will not contribute.  

To complete the evaluation of (\ref{eq:dBk0}), the vertex functions must also be expanded around the point $Q=0$ and  $k_0=E_k$.  This is done using the arguments of the off-shell vertex functions given in Ref.~II.  Expanding these arguments to order $Q^2$, but at order $Q^2$ keeping  only those terms with a factor of $z^2$ (because only they will survive the $z$ integration weighted by $P_2(z)$), gives  
\bea
\tilde R_\pm
&=&k+{\bf R}_\pm 
+(k_0-E_k){\bf S}_\pm
\nonumber\\
R_0^\pm&=&E_k+{\cal E}_\pm +(k_0-E_k) , 
\label{eq:offshellargs}
\eea
 where the small quantities are
\bea
{\bf R}_\pm&=&\pm \frac{zQ }{2m_d}(m_d-E_k)
+\frac{z^2Q^2}{8k\,m^2_d}\Big[k^2-(m_d-E_k)^2\Big]
\nonumber\\
{\bf S}_\pm&=& \mp \frac{zQ}{2m_d}
+ \frac{z^2Q^2}{4k\,m_d^2}(m_d-E_k)
\nonumber\\
{\cal E}_\pm&=&\mp\frac{k_zQ}{2m_d}
\eea
and here it is not necessary to retain any higher powers of $(k_0-E_k)$, because they are multiplied by $Q$ in $\tilde R_\pm$ (and hence are negligible) and are altogether absent from $R_0^\pm$.  Note that  $\tilde R_\pm$ and $R_0^\pm$  reduce to $k$ and $k_0$ at $Q=0$, as expected.  

Expanding the structure functions to  the same order as the expansions (\ref{eq:offshellargs}) gives
%
\bea
h\widetilde X_\pm&\simeq& X + {\bf R}_\pm  X_k +{\cal E}_\pm  X_{k_0}
\nonumber\\&&
+\frac12\Big[ {\bf R}_\pm^2 X_{kk} +2 {\bf R}_\pm {\cal E}_\pm X_{kk_0}+{\cal E}_\pm^2 X_{k_0k_0}\Big] 
 \nonumber\\&&
 +(k_0-E_k)\Big\{X_{k_0}+{\bf S}_\pm X_k+{\bf R}_\pm X_{kk_0}+{\cal E}_\pm X_{k_0k_0}\Big\} 
 \nonumber\\&&
\label{eq:Xexpan}
\eea
%
%
where 
\bea
&&X_k=h\frac{\partial}{\partial k} \widetilde X(k,k_0)\Big|_{Q=0} 
\nonumber\\
 &&X_{k_0}=h\frac{\partial }{\partial k_0} \widetilde X(k,k_0)\Big|_{Q=0}
\eea
 and similarly for the other derivatives.   
 The expansion of the strong form factor will also contribute, and these terms will be discussed separately below.

 It is convenient to express $X_k$ in terms of $X'$, where $X'=h\partial \widetilde X(k,E_k)/(\partial k)$ is the derivative that appears in the calculation of the (A) diagrams.  
Substituting the relations
\bea
X_k&=&X'-\frac{k}{E_k}X_{k_0}
\nonumber\\
X_{kk_0}&=&X'_{k_0}-\frac{k}{E_k}X_{k_0k_0}\nonumber\\
X_{kk}&=&X''-2\frac{k}{E_k}X'_{k_0}-\frac{m^2}{E_k^3}X_{k_0}+\frac{k^2}{E_k^2}X_{k_0k_0}\, ,\qquad
\label{eq:ktoprime}
\eea
where $X'_{k_0}=h\frac{d}{dk}\left(\frac{\partial \widetilde X}{\partial k_0}|_{k_0=E_k}\right)$, into (\ref{eq:Xexpan}) gives  
\bea
&&h\widetilde X_\pm\simeq X \pm\frac{k_zQ}{2km_d}D_{01}(X) +\frac{z^2Q^2}{8k\,m^2_d}D_{02}(X)
\nonumber\\&&\quad
 +(k_0-E_k)\Big\{X_{k_0} \mp\frac{k_zQ}{2km_d}D_{11}(X)+\frac{z^2Q^2}{8k\,m^2_d}D_{12}(X)\Big\},
\nonumber\\&&\label{eq:Xexpan2}
\eea
where the $D_{ij}$'s will be given shortly.  

Calculation of these contributions is very lengthly, and it is therefore useful to estimate the leading terms at the start.  To this end, for the purposes of making estimates only, we recognize that the leading part of the S-state wave function,  $u$, goes like the inverse of the positive energy propagator, which for $k_0\ne E_k$ is
\bea
u(k,k_0)&\sim&\frac{N_0}{\delta_+}=\frac{N_0}{E_k+k_0-m_d}
\nonumber\\
&\to& N_0\left[\frac{k^2}{m}+\epsilon+(k_0-E_k)\right]^{-1}  \, ,
\eea
where $N_0$ is an asymptotic normalization constant and $\epsilon>0$ is the deuteron binding energy.  When $k_0=E_k$ this estimate gives the familiar asymptotic wave function for the deuteron S-state.  From it the size of various derivatives can be estimated:
\bea
u&\sim& ku'\sim k^2 u'' 
\nonumber\\
&\sim& \frac{k^2}{m}u_{k_0}\sim \frac{k^3}{m}u'_{k_0}\sim \frac{k^4}{m^2}u_{k_0k_0}\, . \quad \label{eq:direst}
\eea
This shows that each $k_0$ derivative of the ``positive'' energy wave functions ($u, w$, and $z_\ell ^{-+}$, denoted collectively by $y_+$) is large, of order $m/k$ times larger than each $k$ derivative.  However,  the expressions for the invariants obtained in Ref.~II show that these wave functions are all accompanied by the factor $\delta_+$, and the $k_0$ derivatives of the products $(\delta y)_+\equiv[\delta_+y_+]_{k_0}$ are small corrections, as was shown in the calculation of the magnetic moment presented in Ref.~II.  [Similarily, the ``negative'' energy wave functions ($v_t, v_s$, and $z_\ell^{--}$, denoted collectively by $y_-$) are all accompanied by the factor $\delta_-$, so for these the corresponding derivatives are $(\delta y)_-\equiv[\delta_-y_-]_{k_0}$, and are also small.] 
Since these are small corrections, and the second $k_0$ derivatives are even smaller, we will neglect the second derivatives $[\delta_\pm y_\pm]_{k_0k_0}$.   With these estimates, the $k_0$ derivatives of the wave functions are replaced by
\bea
(y_+)_{k_0}&\to&\frac{m}{k^2}\big[(\delta y)_+-y_+\big]
\nonumber\\
(y_+)_{k_0k_0}&\sim& -\frac{2m}{k^2}(y_+)_{k_0}\to -\frac{2m^2}{k^4}\big[(\delta y)_+-y_+\big] \qquad
\eea  
where, when $k_0=E_k$, $\delta_+=\delta_k\simeq k^2/m$ (neglecting the deuteron binding energy) and $(\delta y)_+$ is a small quantity.

Similar considerations apply to the mixed derivatives, $(y_+)'_{k_0}$.  These are large, but the quantity $(\delta'y)_+\equiv[\delta_+y_+]'_{k_0}$ is small, leading to the following substitution
\bea
(y_+)'_{k_0}\to\frac{m}{k^2}\Big\{(\delta'y)_+-y_+'-\frac2{k}\big[(\delta y)_+-y\big]\Big\}.\quad
\eea
Note that both the second $k_0$ derivatives and mixed derivatives of  $y_+$  generate large contributions to the leading terms involving $y_+$.  Ignoring these contributions will give an incorrect result for the nonrelativistic limit.

With this understanding,  the  $D_{ij}$'s and their leading terms are 
\bea
D_{01}(X)&=&(m_d-E_k )X' - \frac{km_d}{E_k}X_{k_0} \to m X'  - 2kX_{k_0} 
\nonumber\\
D_{02}(X)&=&\Big[k^2-(m_d-E_k)^2\big]\Big]X'+k(m_d-E_k)^2X''
\nonumber\\
&+&\frac{k^3m_d^2}{E_k^2}X_{k_0k_0}
-\frac{2k^2m_d}{E_k}(m_d-E_k)X'_{k_0}
\nonumber\\&&
-\frac{k^3m_d}{E_k^3}(2E_k-m_d)X_{k_0}
\nonumber\\
&\to& -m^2 (X'-k\,X'') -4mk^2X'_{k_0}+4k^3 X_{k_0k_0}
\nonumber
\eea
\bea
D_{11}(X)&=&X'-\frac{k}{E_k}X_{k_0} -(m_d-E_k)X'_{k_0}+\frac{k\,m_d}{E_k}X_{k_0k_0}
\nonumber\\
&&\to X'  -\frac{k}{m} X_{k_0}-mX'_{k_0}+2kX_{k_0k_0}
\nonumber\\
D_{12}(X)&=&2(m_d-E_k)\Big(X'-\frac{k}{E_k}X_{k_0}\Big)
\nonumber\\
&+&\Big[k^2-(m_d-E_k)^2\Big]\Big(X'_{k_0}-\frac{k}{E_k}X_{k_0k_0}\Big)
\nonumber\\&\to&
2(mX' -kX_{k_0})-m(m X'_{k_0}-k X_{k_0k_0})
,\qquad\quad
\nonumber\\ &&
\label{eq:singleDs}
\eea
where the double derivative $X_{k_0k_0}$ 
does not include any of the double $k_0$ derivatives of the $[\delta_\pm y_\pm]$ terms listed above. 

Using this expansion, the generic product of two invariants picks up some cross terms at ${\cal O}(Q^2)$
\bea
h^2X_+Y_-&\simeq& XY+\frac{k_zQ}{2km_d}D_{01}(XY)+\frac{z^2\,Q^2}{8km_d^2}D_{02}(XY)
\nonumber\\&&
+(k_0-E_k)\bigg\{X_{k_0}Y+XY_{k_0}+\frac{k_zQ}{2km_d}D_{11}(XY) 
\nonumber\\&&\qquad\qquad+\frac{z^2Q^2}{8km_d^2}D_{12}(XY)\bigg\}
 ,\qquad\quad\label{eq:XYexpan2}
\eea
where the product coefficients (distinguished from the $D_{ij}(X)$ only by their arguments) are
\begin{widetext}
\bea
D_{01}(XY)&=&D_{01}(X)Y-XD_{01}(Y)
\nonumber\\
D_{02}(XY)&=&D_{02}(X)Y+XD_{02}(Y)-2kD_{01}(X)D_{01}(Y)
\nonumber\\
D_{11}(XY)&=&-D_{11}(X)Y+XD_{11}(Y)+D_{01}(X)Y_{k_0}-X_{k_0}D_{01}(Y)
\nonumber\\
D_{12}(XY)&=&D_{12}(X)Y+XD_{12}(Y)+D_{02}(X)Y_{k_0}+X_{k_0}D_{02}(Y)+2k\big[D_{11}(X)D_{01}(Y)+D_{01}(X)D_{11}(Y)\big] \, ,
\eea
with leading contributions obtained from the leading terms given in (\ref{eq:singleDs}).

Substituting the expansion (\ref{eq:XYexpan2}) into (\ref{eq:dBk0}), taking the differences at $k_0=E_\pm$, and then computing the averages (\ref{eq:QlimB}), gives one set of terms coming from the $k_0$ dependence of the arguments of the invariants,  proportional to the factors $B_{n0,i}^{XY}D_{1(2-n)}^{XY}$, and another coming from the $k_0$ dependence of the expansion coefficients proportional to the factors $B_{n1,i}D_{0(2-n)}^{XY}$.  The generic term is a sum of these two contributions.  Being careful to recall that, through Eq.~(\ref{eq:k0factor}), the factor of $k_0-E_k$ gets converted into the  factor  $-k_zQ/E_k$, and remembering the terms proportional to $B_{02}$ and $B_{03}$ gives 
\bea
\overline{B}_i(k)\Big|_{XY}&=& {\B-}\lim_{Q^2\to0}\frac{m_dm}{Q^2}\frac12\int_{-1}^1dz\Bigg\{\frac{P_2(z)z^2Q^2}{8km_d^2} B_{00,i}^{XY}D_{12}(XY)
+\frac{(k_zQ)^2}{2km_d}B_{10,i}^{XY}(z^2)D_{11}(XY)+Q^2B_{20,i}^{XY}(z^2)(XY)_{k_0}\Bigg\}
\nonumber\\&&
-\lim_{Q^2\to0}\frac{m_dm}{Q^2}\frac12\int_{-1}^1dz\Bigg\{\frac{P_2(z)z^2Q^2}{8km_d^2} B_{01,i}^{XY}D_{02}(XY)  
+\frac{(k_zQ)^2}{2km_d}B_{11,i}^{XY}(z^2)D_{01}(XY)
\nonumber\\&&\qquad\qquad\qquad\qquad
+Q^2\Big[B_{21,i}^{XY}(z^2)+\frac{P_2(z)z^2k^2}{4E_k^4}\Big(\frac12B_{01,i}^{XY}-E_kB^{XY}_{02,i}+E_k^2B^{XY}_{03,i}\Big)\Big]XY\Bigg\}
\nonumber\\
&=&-\frac{m}{60km_d}\Big[B_{00,i}^{XY}D_{12}(XY)+B_{01,i}^{XY}D_{02}(XY)\Big]
-\frac{km}{6}\Big[\overline{B}_{10,i}^{XY}D_{11}(XY)+\overline{B}_{11,i}^{XY}D_{01}(XY)\Big]
\nonumber\\&&\qquad\qquad
-m_dm\Big[\overline{B}_{20,i}^{XY}(X_{k_0}Y+XY_{k_0})+\overline{B}_{21,i}^{XY}\,XY\Big] -\frac{k^2m_dm}{30E_k^4}\Big(\frac12B_{01,i}^{XY}-E_kB^{XY}_{02,i}+E_k^2B^{XY}_{03,i}\Big) XY.\qquad
\label{eq:BXYsub}\qquad
\eea
\end{widetext}
where $(XY)_{k_0}=X_{k_0}Y+XY_{k_0}$, $\overline{B}_{1m,i}^{XY}=B_{1m,i}^{XY}(z^2=\frac35)$ and 
$\overline{B}_{2m,i}^{XY}=B_{2m,i}^{XY}(z^2=\frac13)$.

Now consider the contributions from the $Q$ and $k_0$ dependence of the strong form factors.    Expanding the arguments of the form factors to order $Q^2$ and $(k_0-E_k)^3$ gives
\bea
\tilde p^2 &=& (D_0-k_0)^2-k^2
\nonumber\\
&\to& p^2 -2 (k_0-E_k)(m_d-E_k) + (k_0-E_k)^2
\nonumber\\&&
+\frac{Q^2}{4m_d}\big[ m_d-E_k -(k_0-E_k)\big]\, ,
\eea
with $p^2=m^2+m_d^2-2m_dE_k\simeq m^2-2k^2$.
Hence, the expansion of the form factors can be written
\bea
h^2(\tilde p)&\simeq& h^2+h^2\sum_{nm} Q^n(k_0-E_k)^m B_{nm}^h\, ,
\label{eq:Bexpanofa}
\eea
where $B_{1m}^h=B_{00}^h=B_{22}^h=B_{23}^h=0$ and the exact coefficients, together with their leading values, are
\bea
h^2 B_{01}^h&=&-2(h^2)' (m_d-E_k)\to -h^24m\, a(p^2) 
\nonumber\\
h^2 B_{02}^h&=&(h^2)' + 2(h^2)''(m_d-E_k)^2
\nonumber\\
&\to& h^2[2a(p^2)+4a_2(p^2)]
\nonumber
\eea
\bea
h^2 B_{03}^h&=&-2(h^2)'' (m_d-E_k)- \frac43(h^2)'''(m_d-E_k)^3
\nonumber\\
&\to& -h^2\frac1{m}\Big[4a_2(p^2)+\frac83 a_3(p^2)\Big]
\nonumber\\
h^2 B_{20}^h&=&\frac{(h^2)'}{4m_d}(m_d-E_k)\to h^2 \frac14 a(p^2)
\nonumber\\
h^2 B_{21}^h&=&-\frac{(h^2)'}{4m_d}\to- h^2\frac1{4m}a(p^2)\, ,
\eea
where the derivatives of $h^2$ are with respect to $\tilde p^2$ evaluated at $\tilde p^2=p^2$. 
The first derivative is $(h^2)'=2h^2 a(p^2)$, where $a(p^2)$ was defined previously, and appeared in the discussion of the (A) diagrams.  This definition is generalized  to the higher derivatives
\bea
m^2(h^2)''\equiv 2h^2 a_2(p^2)\qquad m^4(h^2)'''\equiv 2h^2 a_3(p^2)\, ,\qquad
\eea
with $a(p^2)\equiv a_1(p^2)$.  

Using the expansion (\ref{eq:Bexpanofa}) the dependence of the strong form factors can be included by redefining six of the eight expansion coefficients $B_{nm,i}^{XY}$ as follows:
\bea
B_{01,i}^{XY}&\to& B_{01,i}^{XY} + B_{00,i}^{XY} B_{01}^h
\nonumber\\
B_{02,i}^{XY}&\to& B_{02,i}^{XY} + B_{01,i}^{XY} B_{01}^h+B_{00,i}^{XY} B_{02}^h
\nonumber\\
B_{03,i}^{XY}&\to& B_{03,i}^{XY} + B_{02,i}^{XY} B_{01}^h+B_{01,i}^{XY} B_{02}^h+ B_{00,i}^{XY} B_{03}^h
\nonumber\\
B_{11,i}^{XY}&\to& B_{11,i}^{XY} + B_{10,i}^{XY} B_{01}^h
\nonumber\\
B_{20,i}^{XY}&\to& B_{20,i}^{XY} + P_2(z) B_{00,i}^{XY} B_{20}^h
\nonumber\\
B_{21,i}^{XY}&\to& B_{21,i}^{XY} + B_{20,i}^{XY} B_{01}^h + P_2(z) B_{00,i}^{XY} B_{21}^h\, ,
\label{eq: Bh1}
\eea
where care has been taken to  include the factor of $P_2(z)$  from Eq.~(\ref{eq:dBk0}), needed in the last two equations.  Since neither $B_{00,i}^{XY}$ nor the $B^h$ have any $z$ dependence, these $P_2(z)$ terms integrate to zero, and there are no contributions from the $B_{2m}^h$, and $B_{20,i}^{XY}$ is not modified.  

The exact expansion has been retained to this point, but the derivatives of $h^2$ are quite small.  At small $k^2$, $p^2\sim m^2$,  $h=1$, and using the form of $h$ given in Ref.~\cite{Gross:2008ps} (denoted by $H$ in that reference), each successive derivative of $h^2$ is smaller by a factor of $(\Lambda^2-m^2)^{-1}\simeq (2m^2)^{-1}$ (near $k^2\sim0$ and for the values of $\Lambda_N$ found to fit the $np$ data).  Hence successive derivatives of $h^2$ are suppressed by factors of $k^2/m^2$, and to leading order only the first derivative, proportional to $a(p^2)$, need be retained.  With this approximation, and dropping the $P_2(z)$ terms, the relations (\ref{eq: Bh1}) reduce to
\bea
B_{01,i}^{XY}&\to& B_{01,i}^{XY} -4 m\, a(p^2) B_{00,i}^{XY}
\nonumber\\
B_{02,i}^{XY}&\to& B_{02,i}^{XY} -2 a(p^2) (2mB_{01,i}^{XY}-B_{00,i}^{XY})
\nonumber\\
B_{03,i}^{XY}&\to& B_{03,i}^{XY} -2 a(p^2) (2mB_{02,i}^{XY} -B_{01,i}^{XY})
\nonumber\\
B_{11,i}^{XY}&\to& B_{11,i}^{XY} -4 m\, a(p^2) B_{10,i}^{XY} 
\nonumber\\
B_{20,i}^{XY}&\to& B_{20,i}^{XY} 
\nonumber\\
B_{21,i}^{XY}&\to& B_{21,i}^{XY} -4 m\, a(p^2) B_{20,i}^{XY}  \, ,
\label{eq: Bh2}
\eea

Using (\ref{eq:BXYsub}),  $\overline{B}_i(k)$ can be expressed in terms of the invariants $F,G,H,I$ and their first and second derivatives.  These in turn can be written in term of the wave functions $u,w,v_t,v_s$, and $\chi_{\ell}=\{z_0^{--}, z_1^{--},z_0^{-+},z_1^{-+}\}$, the negative $\rho$-spin helicity amplitudes for particle 1, which contribute because the $k_0$ derivatives of the invariants depend on them.   These terms also contributed to the magnetic moment, as discussed in Ref.~I.  The result of the (B) contributions, as introduced in Eq.~(\ref{eq:quadB}) are
\begin{widetext}
\bea
\overline{B}_1(k)&=& 2\pi^2\frac{E_k}{m}\frac{m^2}{10}\Big\{-4\sqrt{2}u'w' -2 w'^2 
-3 v_t'^2+6v_s'^2-\frac{2\sqrt{2}}{k}(6u'w+5v_t'v_s)  -\frac1{k^2}\Big[12 w^2 +35 v_t^2-10\sqrt{2}v_sv_t-48 v_s^2\Big]
\nonumber\\&&\qquad
-\frac{\sqrt{3}\,m}{k}\Big[w'v_t'-3\frac{w'v_t}{k}
-10\frac{w v_t}{k^2}\Big]  
+\frac{2m}{k}\Big[3v_s' z_\delta'+\frac{1}{k}v_s'z_\delta+\frac{14}{k^2}v_sz_\delta\Big] +B_{1{\rm D}}+2a(p^2) B_1^h + \Delta B_1\Big\}
\nonumber\\
\overline{B}_2(k)&=& 2\pi^2\frac{E_k}{m}\frac{m^2}{10}\Big\{-20\sqrt{2}\,\frac{v_t'v_s}{k} -\frac1{k^2}\Big[30 v_t^2-20\sqrt{2}\,v_tv_s\Big] + \Delta B_2\Big\} \label{eq:BQterms}
\eea
where the D-type corrections are
\bea
B_{1{\rm D}}&=&3\Big\{\sqrt{2}\big(u' [\delta_+\hat w]_{k_0}'+w'[\delta_+\hat u]_{k_0}'\big)+w'[\delta_+\hat w]_{k_0}'+v_t' [\delta_-\hat v_t]_{k_0}' -2v_s' [\delta_-\hat v_s]_{k_0}'\Big\} 
\nonumber\\
&&+ \frac2{k}\Big[(9\sqrt{2}u'+5w')[\delta_+\hat w]_{k_0}-5\sqrt{2} \,w'[\delta_+\hat u]_{k_0} +2 v_t'[\delta_-\hat v_t]_{k_0}-4 v_s'[\delta_-\hat v_s]_{k_0}\Big]
\nonumber\\&&
+\frac2{k^2}\Big[(10\sqrt{2}\,u+27 w-2[\delta_+\hat w]_{k_0})[\delta_+\hat w]_{k_0}-\sqrt{2}(w+4[\delta_+\hat w]_{k_0})[\delta_+\hat u]_{k_0}+7v_t[\delta_-\hat v_t]_{k_0}-14v_s[\delta_-\hat v_s]_{k_0}\Big]\, ,\qquad \label{eq:B1D}
\eea
and we introduced the difference 
\bea
z_\delta=\sqrt{2}z_0^{--}-z_1^{--}. \label{eq:zdelta}
\eea
The $m/k$ terms were reduced using the identities
\bea
\int_0^\infty dk\frac{d}{dk}(kz_1z_2')&=& \int_0^\infty k^2 dk \Big[\frac{1}{k}(z_1'z_2'+z_1z_2'') +\frac{z_1z_2'}{k^2} \Big]=0  
\nonumber
\eea
\bea
\int_0^\infty dk\frac{d}{dk}(z_1z_2)&=&\int_0^\infty k^2 dk\Big(\frac{z_1'z_2+z_1z_2'}{k^2}\Big)=0\, .
\nonumber\\&&
\eea
The leading contributions from the derivatives of $h^2$ are
\bea
B_1^h&=&-2k^2\Big\{4\sqrt{2}u'w'+2w'^2+\frac{12\sqrt{2}}{k}u'w
+\frac1{k^2}(\sqrt{2}uw+6w^2)\Big\}+\frac{m}{k}16\sqrt{3}\,wv_t
-8m^2\Big\{v_t'^2-2v_s'^2 +\frac2{k^2}(v_t^2-2v_s^2)\Big\}\, ,
\nonumber\\&&
\eea
\end{widetext}
and the $k^2/m^2$ corrections to the leading terms are
\bea
\Delta B_1&=&-\frac{k^2}{m^2}\Big\{4\sqrt{2}u'w'+2w'^2+\frac{19\sqrt{2}}{k}u'w
\nonumber\\&&
-\frac1{2k^2}(27\sqrt{2}\,uw-88w^2)\Big\}
\nonumber\\
\Delta B_2&=&\frac{3}{m^2}\Big\{\sqrt{2}uw-w^2\Big\}\, .
\eea

\subsubsection{Evaluation of the regular terms}

The contributions from the ${\cal C}$ traces are finite, and the generic term from $\delta{\cal C}_ i$ that contributes to the quadrupole moment has the form
\bea
\delta{\cal C}_i\big|_{XK}&=&\big[P_2(z)C_{0,i}^{XZ}+k_z Q\, C^{XK}_{1,i} +Q^2 C^{XK}_{2,i}\big]h^2\tilde X_+\tilde K_- 
\nonumber\\&& \label{eq:dC}
\eea
where $C_{1,i}$ is linear and $C_{2,i}$ quadratic in $k_z^2$.   
The contributions from the first term come from the expansion of the arguments of the wave functions to order $Q^2$ (but, because of the presence of $P_2(z)$, only coefficients proportional to $z^2$ will contribute) and from the second term to order $Q$.   Expanding the arguments given in Ref.~II up to order $Q^2$ gives
\bea
R_+&\simeq& k -\frac{zQ}{2m_d}E_k-\frac{z^2 Q^2 m^2}{8km_d^2}
\nonumber\\
\hat R_-&\simeq& k -\frac{zQ}{2m_d}m_\Delta
-\frac{z^2Q^2}{8km_d^2}\big[m_\Delta^2-k^2\big]
\nonumber\\
\hat R^-_0&\simeq& E_k+\frac{k_zQ}{2m_d}
\eea
where we have introduced $m_\Delta\equiv 2 m_d-E_k$, and  $R_+$ is the argument of the final on-shell vertex function invariants $X_+$, and $\hat R_-$ and $\hat R^-_0$ the arguments of the initial $K_-$ invariants with both particles off-shell.   
Hence, expanding  a typical product of vertex invariants to order $Q^2$ gives
\bea
h^2\tilde X_+\tilde K_- &\simeq& XK
- \frac{zQ}{2m_d}D_1(XK) -\frac{z^2Q^2}{8km_d^2}D_2(XK)
\nonumber\\&&
\eea
where 
\begin{widetext}
\bea
D_1(XK)&=&E_k X'K+m_\Delta XK_{k}-k XK_{k_0}
=E_kX'K+m_\Delta XK'-\frac{2 k m_d}{E_k}XK_{k_0}
\nonumber\\
&\to& m(X'K+3XK')-4k X K_{k_0}
\nonumber\\
D_2(XK)&=&m^2X'K+\big[m_\Delta^2-k^2\big]XK_k 
-kE_k^2X''K-k\,m_\Delta^2XK_{kk} -k^3  XK_{k_0k_0} +2k^2m_\Delta XK_{kk_0}  -2kE_km_\Delta X'K_k 
\nonumber\\
&&  +2k^2E_k X'K_{k_0}
=m^2X'K+\big[m_\Delta^2-k^2\big]XK' -kE_k^2X''K -k\,m_\Delta^2XK''
 -\frac{4k^3 m_d^2}{E_k^2} XK_{k_0k_0}
\nonumber\\&& +\frac{4k^2 m_d m_\Delta}{E_k} XK'_{k_0}-2kE_km_\Delta X'K'
+4k^2m_d X'K_{k_0}-\frac{4k^3m_d}{E_k^3}(m_d-E_k)XK_{k_0}
\nonumber\\
&\to& m^2(X'K+9XK'-kX''K-9kXK''-6kX'K')-16 k^3XK_{k_0k_0} +8k^2m(3XK'_{k_0}+X'K_{k_0}).
\eea
These were transformed using (\ref{eq:ktoprime}) before the leading terms were extracted.  Hence the contributions to the quadrupole moment coming from the ${\cal C}$ traces are of the form
\bea
\overline C_i(k)=\frac1{m}(\overline{\cal C}_{i+}+{\cal C}_{i-})&=&\lim_{Q^2\to0}\frac{m_d}{mQ^2}\int_{-1}^1dz\Big\{-\frac{z^2P_2(z)Q^2}{8km_d^2}C_{0,i}^{XK}D_2(XK)-\frac{z^2kQ^2}{2m_d}C^{XK}_{1i}D_1(XK)+Q^2C^{XK}_{2,i}XK\Big\}
\nonumber\\
&=&-\frac{1}{30kmm_d}{C}_{0,i}^{XK}D_2(XK)-\frac{k}{3m}\overline{C}^{XK}_{1i}D_1(XK)+\frac{2m_d}{m}\overline{C}^{XK}_{2,i}XK\, ,
\eea
where $\overline{C}^{XK}_{1i}=C^{XK}_{1i}(z^2=\frac35)$, $\overline{C}^{XK}_{2i}=C^{XK}_{2i}(z^2=\frac13, z^4=\frac15)$ and we used the fact that $\overline{\cal C}_{i+} =\overline{\cal C}_{i-}$.

To include the contributions from the derivatives of the strong form factor,  $h_+=h(p_+)$ (where $p_+^2=m_d^2+m^2-2D_0E_k+Qk_z$), expand to order $Q^2$, giving
%
\bea
h_+^2\simeq  h^2+2h^2a(p^2)  \Big(Qk_z-\frac{Q^2}{4m_d}E_k\Big)\, .
\eea
Because the $Q^2$ term includes no $z$ dependence, it will make no contribution, and the effect of the linear term is to modify the $C_{2,1}^{XK}$ of Eq.~(\ref{eq:dC}) by adding a term
\bea
C_{2,1}^{XK}\to C_{2,1}^{XK}+2a(p^2)k_z^2C_{1,i}^{XK}\, .
\eea
However, there are no leading contributions from these terms.

The leading contribution to the quadrupole moment coming from the $\cal{C}$ traces are therefore
%
\bea
\overline{\cal C}_1(k)&=&2\pi^2 \frac{E_k}{m}\frac{m^2}{10}\Big\{12\sqrt{2} v_t'v_s'+\frac{92\sqrt{2}}{k} v_t'v_s-\frac1{4k^2}(15 v_t^2+68\sqrt{2}v_tv_s)+\frac{m}{k}\Big[20v_s'z_\delta'-\frac{122}{k}v_s'z_\delta-\frac{164}{k^2}v_s z_\delta\Big]+\Delta C_1\Big\}\qquad
\nonumber\\
\overline{\cal C}_2(k)&=&2\pi^2 \frac{E_k}{m}\frac{m^2}{10}\Big\{-20\sqrt{2}\,\frac{v_t'v_s}{k}-\frac{1}{k^2}(30v_t^2-20\sqrt{2}\,v_tv_s)+\Delta C_2\Big\} \label{eq:CQterms}
\eea
where $z_\delta$ was defined in Eq.~(\ref{eq:zdelta}).  Note that these terms all depend the P-state components, but that they have very large coefficients.  The $k^2/m^2$ corrections from the large components are
\bea
\Delta C_1&=&\frac12\Delta C_2=-\frac{5k}{m^2}\Big[\sqrt{2}\,u'w+\frac1{2k}(3\sqrt{2}\,uw+w^2)\Big]
\eea

Finally, the combined contribution to the quadrupole moment from the  (B)+B$_\pm$ terms is the sum of the terms from (\ref{eq:BQterms}) and (\ref{eq:CQterms})
\bea
Q_B&=&e_0\frac{m^2}{10}\int_0^\infty k^2dk\Big\{-(4\sqrt{2}u'w' +2 w'^2)\Big[1+\frac{k^2}{m^2}\Big]  +12\sqrt{2}v_t'v_s'
-3 v_t'^2+6v_s'^2-\frac{\sqrt{2}}{k}(12u'w-82v_t'v_s)  
\nonumber\\&&\qquad
-\frac1{k^2}\Big[12 w^2 +50 v_t^2+58\sqrt{2}v_sv_t-48 v_s^2\Big] -\frac{\sqrt{3}\,m}{k}\Big[w'v_t'-3\frac{w'v_t}{k}
-10\frac{w v_t}{k^2}\Big]
\nonumber\\&&\qquad
+\frac{m}{k}\Big[26v_s' z_\delta'-\frac{120}{k}v_s'z_\delta-\frac{136}{k^2}v_sz_\delta\Big] +B_{1{\rm D}}+2a(p^2) B^h_1
-\kappa_s\Big[40\sqrt{2}\,\frac{v_t'v_s}{k}+\frac{1}{k^2}(60v_t^2-40\sqrt{2}\,v_tv_s)\Big]
\nonumber\\&&\qquad
-\frac{1}{m^2}\Big[\sqrt{2}\,k\, u'w(24+10\kappa_s)+\sqrt{2}\,uw(12-6\kappa_s)+\frac12 w^2(93+16\kappa_s)
\Big]\Big\}\, .  \qquad\label{eq:QBterms}
\eea

\subsection{Total contribution}

Adding the contributions from (\ref{eq:A12lead}), (\ref{eq:QAaterms}), (\ref{eq:Qhprime}), and (\ref{eq:QBterms}), and setting $2e_0=1$, gives the leading result for the quadrupole moment as the sum of eight terms.  Dividing by $m_d^2\simeq 4m^2$ gives 
\bea
Q_d=Q_{\rm NR}+Q_{Rc}+Q_{h'}
+Q_{V_2}+Q_{V_1}+Q_{\rm int}+Q_{P}+Q_\chi
\eea
where these terms are
\bea
Q_{\rm NR}&=&-\frac{1}{40}\int_0^\infty k^2\,dk\Big\{4\sqrt{2}\,u'w'+2w'^2+\frac{12\sqrt{2}}{k}u'w+\frac{12}{k^2}w^2\Big\}=\frac{\sqrt{2}}{10}\int_0^\infty r^2dr\Big\{uw-\frac{w^2}{\sqrt{8}}\Big\}
\nonumber\\
Q_{Rc}&=&\frac1{80} \int_0^\infty \frac{k^4}{m^2}\,dk\Big\{\frac{\sqrt{2}}{k} u'w(2\kappa_s-18)+\frac{6\sqrt{2}}{k^2}uw(1-\kappa_s)-\frac1{2k^2}w^2(87+4\kappa_s)\Big\}\nonumber\\
Q_{h'}&=&\frac{1}{80}\int_0^\infty k^2\,dk \,2 a(p^2)\bigg\{-2k^2\Big[8\sqrt{2}u'w'+4w'^2+\frac{24\sqrt{2}}{k}u' w+\frac{1}{k^2}(\sqrt{2}uw+12w^2)\Big]+16\sqrt{3}\frac{m}{k}wv_t 
\nonumber\\&&\qquad
-16m^2\Big[v_t'^2-2v_s'^2+\frac{2}{k^2}(v_t^2-2v_s^2)\Big]\bigg\}
\nonumber\\
Q_{V_2}&=&\frac{1}{20}\int_0^\infty k^2\,dk\bigg\{\sqrt{2}\,(u'{w^{(2)}}'+{u^{(2)}}'w')+w'{w^{(2)}}'-v_t'{v_t^{(2)}}'+2v_s'{v_s^{(2)}}' +\frac{3\sqrt{2}}{k}({u^{(2)}}'w +u' w^{(2)})
\nonumber\\&&\qquad
+\frac1{k^2}\Big[6ww^{(2)}-2 v_tv_t^{(2)}+4v_sv_s^{(2)}\Big] \bigg\}
\nonumber\\
Q_{V_1}&=&\frac{1}{80}\int_0^\infty k^2\,dk \, B_{1D}
\nonumber\\
Q_{\rm int}&=& -\frac{\sqrt{3}}{80}\int_0^\infty k^2\,dk  \, \frac{m}{k}\Big[w'v_t'-3\frac{w'v_t}{k}
-10\frac{w v_t}{k^2}\Big] 
\nonumber\\
Q_{P}&=&\frac{1}{80}\int_0^\infty k^2\,dk \,\bigg\{2v_s'^2-v_t'^2+12\sqrt{2} v_t'v_s'+(82-40\kappa_s)\frac{\sqrt{2}}{k}v_t'v_s -\frac2{k^2}[(23+30\kappa_s)v_t^2+(29-20\kappa_s)\sqrt{2}v_tv_s-20v_s^2] \bigg\}
\nonumber\\
Q_{\chi}&=& \frac{1}{40}\int_0^\infty k^2\,dk\, \frac{m}{k} \Big[13v_s' z_\delta'-\frac{60}{k}v_s'z_\delta-\frac{68}{k^2}v_sz_\delta\Big]  \label{eq:result}
\eea
where $B_{1D}$ was given in Eq.~(\ref{eq:B1D}).

\end{widetext}

\end{document}